\documentclass[10pt,amsmath,amssymb,nofootinbib,twoside,twocolumn,superscriptaddress,floats,floatfix,aps,prd,preprintnumbers]{revtex4-2}
\usepackage[british]{babel}
\usepackage{xcolor}
\usepackage{booktabs}
\usepackage{latexsym}
\usepackage{amssymb}
\usepackage{amsmath}
\usepackage{graphicx}
\usepackage{tabularx}
\usepackage{cancel}
\usepackage{hyperref}
\usepackage{bm}
\usepackage{aas_macros}
\hypersetup{
    colorlinks=true,
    linkcolor=blue,
    citecolor=blue,
    filecolor=black,
    urlcolor=blue,
    pdfauthor={Ivan~De~Martino},
    pdftitle={XXX}
}
\usepackage[capitalise]{cleveref}
\usepackage[utf8]{inputenc}
\usepackage[normalem]{ulem}
\usepackage{bm}
\begin{document}
\newcommand{\idm}[1]{{\color{red} \bf{#1}}}
\newcommand{\rdm}[1]{{\color{blue} {#1}}}

\newcommand{\Salamanca}{\affiliation{Departamento de F\'isica Fundamental, Universidad de Salamanca, Plaza de la Merced, s/n, E-37008 Salamanca, Spain}}
\newcommand{\SalamancaIUFFyM}{\affiliation{Insituto Universitario de F\'isica Fundamental y Matem\'aticas (IUFFyM), Universidad de Salamanca, Plaza de la Merced, s/n, E-37008 Salamanca, Spain}}
\newcommand{\UCM}{\affiliation{Departamento de Física Teórica and IPARCOS, Universidad Complutense de Madrid, E-28040 Madrid, Spain}}

\author{Ivan~De~Martino}
\email{ivan.demartino@usal.es}
\Salamanca
\SalamancaIUFFyM

\author{Riccardo~Della~Monica}
\email{rdellamonica@usal.es}
\Salamanca

\author{Diego Rubiera-Garcia}
\email{drubiera@ucm.es}
\UCM

\title{Constraints on metric-affine gravity black holes from the stellar motion at the Galactic Center}

\begin{abstract}
    We consider a static, spherically symmetric space-time with an electric field arising from a quadratic metric-affine extension of General Relativity. Such a space-time is free of singularities in the centre of the black holes, while at large distances it quickly boils down to the usual Reissner-Nordstr\"om solution. We probe this space-time metric, which is uniquely characterized by two length scales, $r_q$ and $\ell$,  using the astrometric and spectroscopic measurements of the orbital motion of the S2 star around the Galactic Center. Our analysis constrains $r_q$ to be below $2.7M$ for values $\ell<120 AU$, strongly favouring a central object that resembles a Schwarzschild black hole. 
\end{abstract}

\maketitle


\section{Introduction}

Intense research activity over the last three decades has achieved unprecedented results in black hole physics. Both technological and theoretical advancements have led to the long-awaited first gravitational wave detection of a binary black hole merger event in 2016 by the LIGO/VIRGO collaboration \cite{LIGOScientific:2016aoc} and to the challenging direct imaging of the accreted plasma in the vicinity of the event horizon of the supermassive black holes M87* (at the centre of the M87 galaxy) in 2019 \cite{EventHorizonTelescope:2019dse} and Sagittarius A* (SgrA*, in the Galactic Center of our Galaxy) in 2022 \cite{EventHorizonTelescope:2022wkp} by the Event Horizon Telescope collaboration. Along with these direct observational pieces of evidence, other indirect probes have been gathered over the years that provide additional support to the black hole paradigm (i.e. the universality of the Kerr solution characterized solely by mass and angular momentum to describe every black hole in the Universe) as is described by General Relativity (GR). A remarkable example is the observation of the S-stars cluster in the Galactic Center of the Milky Way \cite{deLaurentis:2022oqa}. Such stars are accelerated at very high orbital velocities ($\sim 2.5\%$ of the speed of light) by a point-like gravitational source located exactly in the SgrA* region, which provided early evidence of a supermassive compact object in this region \cite{Eckart:1996zz, Ghez:2003rt}. The orbital tracking of these stars over the past 30 years has allowed to derive an increasingly precise estimate of the mass of SgrA*, $M\sim4.2\times10^6\,M_\odot$, and of its distance from us, $D\sim 8$ kpc, and has recently allowed the detection of relativistic effects on the orbit of the brightest star in the cluster, S2 \cite{GRAVITY:2018ofz, Do:2019txf, GRAVITY:2020gka}. 

The bulk of all the observational evidence that we nowadays possess in favour of the existence of black holes is remarkable. With the increasing precision of astronomical observations, the margin of possible deviations from the standard general relativistic description of these objects is getting narrower and narrower. This inevitably clashes with one of the most outstanding unresolved theoretical flaws of GR, \emph{i.e.} the existence of space-time singularities at the center of black holes, where classical determinism is lost and the theory itself breaks apart \cite{Senovilla:2014gza}. The existence of singularities in GR is an unavoidable consequence whenever there exists a future trapped surface, the matter energy-momentum tensor satisfies the null energy condition, and global hyperbolicity is fulfilled \cite{Penrose:1964wq, Hawking1966, Tipler1977}. Singularities are commonly related to the divergence of geometric invariants (scalars), constructed from the Riemann tensor, yet the theorems on singularities relate them instead to the incompleteness of geodesic trajectories within the space-time \cite{Ellis1977}. Different avenues have been investigated to formulate black hole solutions that avoid generating singularities. A typical avenue is to relax the null energy condition, introducing exotic forms of matter-energy sources, and leading to both alternative black holes and horizonless compact objects \cite{Bardeen1968, AyonBeato1998, Bronnikov2006, Bambi2023}. From a different perspective, the typical occurrence of the divergence of some sets of curvature scalars as the singularity is approached by some causal trajectories might be interpreted as the need to supersede GR at the Planck scale with new gravitational corrections to make it compatible with quantum mechanics, \emph{i.e.}, to formulate a quantum theory of gravity. Since the latter is not yet at hand (if it will ever be), one can resort to the so-called modified theories of gravity, that is, extensions of GR via different recipes, see e.g. \cite{Sotiriou:2008rp, Clifton:2011jh, Nojiri:2017ncd} for some reviews.

In this work, we take the latter path and consider a spherically symmetric, electrically charged system coupled to a metric-affine extension of GR. Such an extension adds quadratic terms in curvature to the Einstein-Hilbert action of GR and contains families of solutions that are free of incomplete geodesic trajectories, and in some cases also provide finite sets of curvature scalars everywhere \cite{Olmo2012a, Olmo2012b, Olmo2012c, Olmo:2015bya, Olmo2022}. The key to these results seems to lie in the metric-affine (or Palatini) formulation of the theory \cite{Olmo:2011uz}, since it maintains the second-order character of the field equations without incurring in the generation of the ghost-like instabilities of its metric cousin\footnote{In the Riemannian (metric) formalism, the connection is metric-compatible a priori, \emph{i.e.}, it is given by the Christoffel symbols of the metric. In the Palatini formalism, on the other hand, this assumption is relaxed and the affine connection is determined through the field equations, which have thus a fixed second order.}. In turn, this formulation brings to exact analytical solutions, first derived in \cite{Olmo2012a}, and which are characterized by the asymptotic mass of the central object, $M$, its charge ``length", $r_q^2=2G_N q^2$, and a new length scale $\ell$, the latter associated to the higher-order curvature corrections in the gravity Lagrangian.

The main aim of this work is to determine the observational viability of the family of metric-affine black hole solutions mentioned above using to this end the motion of the S2 star in the Galactic Center, taking advantage of publicly available data to derive constraints on $r_q$ and $\ell$ at such scales. To do so, we shall analyze time-like geodesics in these space-times, describing the trajectories of massive-test particles around the central supermassive object of SgrA*. This work is organized as follows. In Section \ref{sec:theory} we provide a thorough description of the family of space-times that we aim to constrain; in Section \ref{sec:numerical} we explain in detail our orbital model, the parameters it involves, and the numerical procedure that we have developed to derive orbits for S2 in this model; in Section \ref{sec:data} we report details on the public datasets that we have used and on our statistical analysis; in Section \ref{sec:results} we show the results of our Bayesian analysis; finally, Section \ref{sec:conclusions} is devoted to conclusions and final remarks.

\section{Theoretical background}
\label{sec:theory}

\subsection{The theory of gravity and matter}

In the standard metric formulation of classical gravitation, the metric is the sole character while the connection is regarded as a secondary object and set as the Levi-Civita one, namely, it is given by the Christoffel symbols of the metric. As opposed to that, in the metric-affine (Palatini) formulation, metric and connection are independent entities, to be determined via variations of the action of the theory with respect to each of them \cite{Olmo:2011uz}. For a large class of such metric-affine theories which include in their definitions only contractions with the symmetric part of the Ricci tensor (dubbed as Ricci-based gravities \cite{Afonso:2018bpv}), the field equations turn out to be second order and no ghost-like propagating degrees of freedom are present (for a detailed explanation of the reason why see Ref. \cite{BeltranJimenez:2019acz}). For the sake of this work, we shall consider a quadratic extension of GR coupled to a sourceless electric field, described by the action 
\begin{eqnarray}\label{eq:action}
    \mathcal{S}&=&\frac{1}{2\kappa^2} \int d^4x \sqrt{-g} \left[ R + \ell^2 (aR^2 + R_{\mu\nu}R^{\mu\nu}) \right]  \nonumber \\
    &-&\frac{1}{16\pi} \int d^4x \sqrt{-g} F_{\mu\nu}F^{\mu\nu}.
\end{eqnarray}
Here, $\kappa^2\equiv 8\pi G$ is the usual GR gravitational constant, $a$ is a dimensionless constant, $g$ is the determinant of the space-time metric $g_{\mu\nu}$, and $\ell$ (with dimensions of a length) modulates the higher-curvature corrections in the Lagrangian. The quantities $R=g^{\mu\nu}R_{\mu\nu}(\Gamma)$ and $R_{\mu\nu}(\Gamma)$ are respectively the Ricci scalar and the Ricci tensor related to the affine connection $\Gamma \equiv \Gamma_{\mu\nu}^{\lambda}$, which we recall is independent of the space-time metric $g_{\mu\nu}$. As for the matter fields, we have introduced the Maxwell tensor $F_{\mu\nu}=\partial_{\mu}A_{\nu}-\partial_{\nu}A_{\mu}$ related to the 4-vector potential $A_{\mu}$.

The choice of the action \eqref{eq:action} is justified on the grounds of results from the theory of quantized fields in curved space-times \cite{Parker2009}, which requires the introduction of higher-order curvature corrections suppressed by a (length-squared) scale, the latter typically associated to Planck-scale effects. In this work, however, we shall take $\ell$ as a free parameter, to be observationally constrained. 

\subsection{The family of solutions}

In this context, exact analytical solutions for the space-time metric associated with such a theory have been derived in the literature and analyzed in detail (see \cite{Olmo2012a, Olmo2012b, Olmo2012c} for more details\footnote{It is worth mentioning that, due to a happy coincidence explained in \cite{Afonso:2018mxn}, these are also solutions of another metric-affine theory of gravity, the so-called Eddington-inspired Born-Infeld one \cite{Olmo:2013gqa}.}). In ingoing Eddington-Finkelstein coordinates and considering geometrized units, $G_N = c = 1$, the corresponding line element can be written as 
\begin{equation}
    \label{eq:metric_EF}
    ds^2=-A(x)dv^2+\frac{2}{\sigma_+}dvdx+r^2(x)d\Omega^2 \ ,
\end{equation}
with the following definitions
\begin{eqnarray}\label{eq:A}
    A(x)&=& \frac{1}{\sigma_+}\left[1-\frac{r_s}{ r  }\frac{(1+\delta_1 G(r))}{\sigma_-^{1/2}}\right], \\
    \delta_1&=& \frac{1}{2r_s}\sqrt{\frac{r_q^3}{\ell}}, \\
    \sigma_\pm&=&1\pm \frac{r_c^4}{r^4(x)}, \\
    r^2(x)&=& \frac{x^2+\sqrt{x^4+4r_c^4}}{2} \label{eq:r(x)}.
\end{eqnarray}
Here $r_s = 2M$ is the Schwarzschild radius of the central object, with $M$ its mass as seen from an asymptotic observer; the quantity $r_c=\sqrt{\ell r_q}$ is a combination of the characteristic length $\ell$ of the theory and of the charge-related length $r_q^2=2G_N q^2$. As for the function $G(z)$ characterizing the metric function, with $z=r/r_c$ a dimensionless radial coordinate, it is given by
\begin{equation}
    G(z)=-\frac{1}{\delta_c}+\frac{1}{2}\sqrt{z^4-1}\left[f_{3/4}(z)+f_{7/4}(z)\right] \ ,
    \label{eq:G(z)}
\end{equation}
where $\delta_c\approx 0.572069$ is a constant and $f_\lambda(z)={_2}F_1 \left[\frac{1}{2},\lambda;\frac{3}{2};1-z^4\right]$ with ${_2}F_1(a,b;c;z)$ the gaussian hypergeometric function. 

The above line element is a generalization of the Reissner-Nordst\"om (RN) geometry of GR. Indeed, for large values of $z$ (corresponding to the limit $r\gg r_c$), the function $G(z)$ in Eq. \eqref{eq:G(z)} tends to $G(z)\sim -1/z$. In the same limit, $\sigma_\pm \to 1$, bringing the two radial coordinates to approach each other, $r^2(x) \to x^2$, and then in such a limit the metric function boils down to
\begin{equation} \label{eq:metric_RN}
    A(x)\approx 1-\frac{r_s}{ r  }+\frac{r_q^2}{2r^2} \ ,
\end{equation}
which is the usual RN geometry of GR expressed in terms of our variables.

Redefining the time coordinate in  Eq.\eqref{eq:metric_EF} according to $dv= dt+dx/(A\sigma_+)$, yields a different expression for the line element that resembles more closely the RN solution in GR
\begin{equation}\label{eq:metric}
    ds^2=-A(x)dt^2+\frac{1}{B(x)}dx^2+r^2(x)d\Omega^2 \ ,
\end{equation}
where $B(x)=A(x)\sigma_+^2$. To preserve the simple representation of $r^2(x)$ introduced in Eq.\eqref{eq:r(x)}, we avoid absorbing the factor $\sigma_+$ into a redefinition of the coordinate $x$. Such a radial coordinate, $x$, is defined on the whole real line, \emph{i.e.} $x\in ]-\infty,+\infty[$, which brings the quantity $r(x)$ to have a minimum $r_c$ in correspondence with $x=0$. This fact implies that the surface area of hypersurfaces of constant $t$ and $x$ (which has the geometry of a 2-sphere with radius $r(x)$), given by $S=4\pi r^2(x)$, possesses a minimum for $x=0$. This distinctive feature, \emph{i.e.} a bounce in the (radial coordinate of the) geometry, indicates a non-trivial topological structure in the form of a wormhole \cite{Visser1995}, with $r=r_c$ identified as its throat. Moreover, since $dx^2=\sigma_+^2 dr^2/\sigma_-$, at $x=0$ (\emph{i.e.} at $r=r_c$), one finds $dr/dx=0$ there. This readily implies that a coordinate transformation from $x$ to $r(x)$ is ill-defined at the throat. For this reason, if one wishes to use $r$ as the radial coordinate, one has to restrict intervals in which $r(x)$ is a monotonic function \cite{Stephani2009}. This requires splitting the domain of $r$ into two regions, $x>0$, and $x<0$; however, the transition from one region to the other is smooth, and allows the completion of all geodesic trajectories across $x=0$ \cite{Olmo:2015bya}. Furthermore, the presence of potential curvature divergences have been shown in  \cite{Olmo:2016fuc} to not exert any utterly destructive process upon extended (time-like) observers. These features allow to consistently interpret these solutions as regular black holes and naked geometries, depending on the interplay of parameters (see below). Nonetheless, for the sake of this work, we find it more convenient to use $x>0$ as the radial coordinate, since we are only interested on the geometry of that side of the throat.

\subsection{The classes of configurations}

A characterization of the causal structure of the geometry in terms of the values of the free parameters that appear in Eq. \eqref{eq:metric_EF} is possible by considering that close to the central object, \emph{i.e.} for $r\to r_c$ (or equivalently $x\to 0$), important departures from the RN solution are present. If we now consider a power series expansion of $A(r(x))$ around $x=0$, we get
\begin{eqnarray}\label{eq:A_expansion}
    \lim_{r\to r_c} A(x) &\approx & \frac{N_q}{4N_c}\frac{\left(\delta _1-\delta _c\right) }{\delta _1 \delta _c }\sqrt{\frac{r_c}{ r-r_c} }+\frac{N_c-N_q}{2 N_c} \nonumber \\
&+&O\left(\sqrt{r-r_c}\right) \ ,
\end{eqnarray}
where we have defined the number of charges as $N_q=q/e$ (being $e$ the electron charge) and $N_c\equiv \sqrt{2/\alpha}\approx 16.55$ (being $\alpha$ the fine structure constant). One can then show that the causal structure of the space-time in Eq. \eqref{eq:metric} depends on the value of the charge-to-mass ratio, $\delta_1$, with respect to the critical value $\delta_c$ \cite{Olmo2012a, Olmo2012b, Olmo2012c}. Namely:

\begin{itemize}
\item For $\delta_1<\delta_c$ an event horizon always exists on both sides of the throat, regardless of the value of $N_q$. Due to the existence of the horizons, this can be regarded as a black hole solution, close to the Schwarzschild one, from the viewpoint of observers located on either side of the throat, outside the respective horizon.
\item For $\delta_1>\delta_c$, on the other hand, the space-time causal structure is more complicated and presents either two (non-degenerate) horizons, a degenerate one, or no horizons at all, on each side of the throat, depending on the value of $N_q$ (see \cite{Olmo2012a} for a thorough analysis of the horizons and causal structure this case). In all cases, however, the geometry of space-time towards the centre exhibits drastic changes with respect to its GR counterpart.
\item In the extremal case $\delta_1=\delta_c$, we have three sub-cases depending on the value of $N_q$. In particular, if  $N_q>N_c$, there are two horizons symmetrically located with respect to the throat; in the case  $N_q= N_c$, the two horizons coincide at the throat, $r=r_c$ (or $x=0$); finally if $N_q<N_c$ the horizons disappear bringing a sort of black hole remnant that can be regarded as the endpoint of Hawking evaporation or might be generated by large density fluctuations in the primordial Universe \cite{Hawking1971}. For all these configurations curvature scalars are everywhere finite \cite{Olmo:2016fuc}. 
\end{itemize}

\begin{figure}
    \includegraphics[width = 0.75\columnwidth]{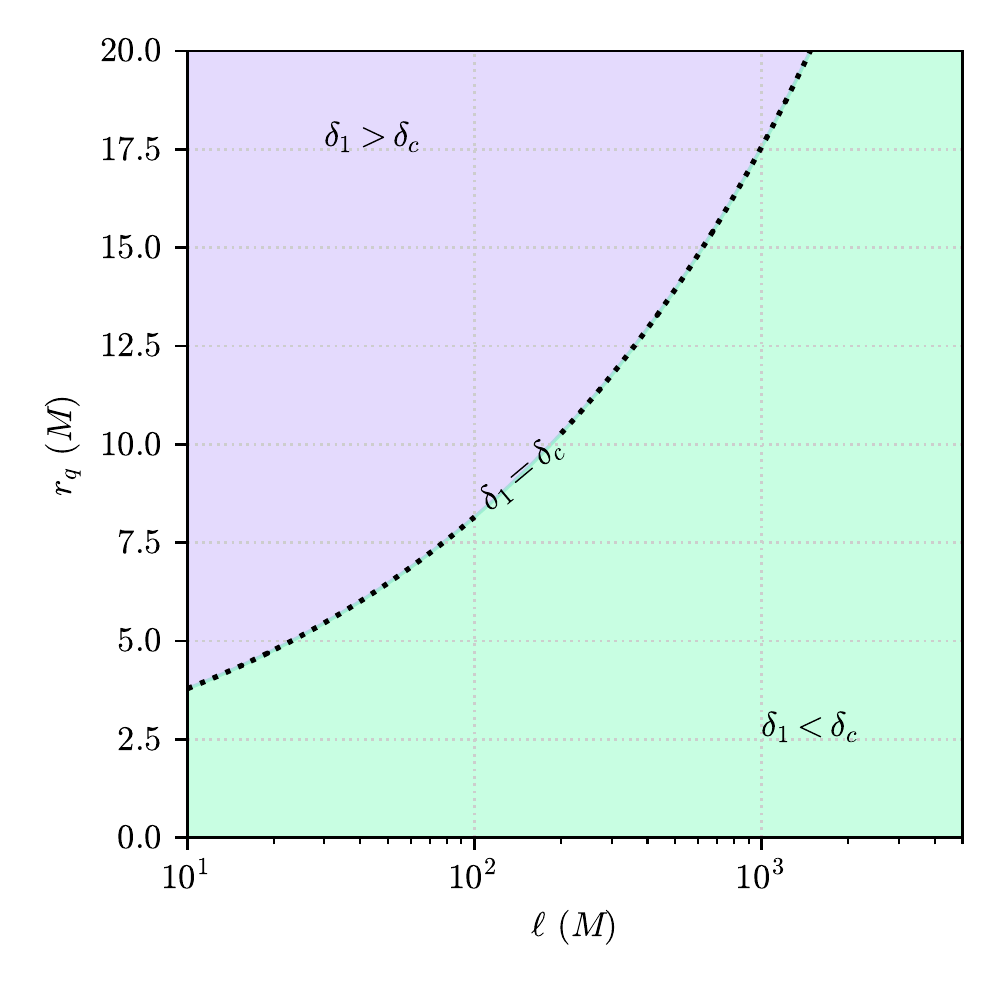}
    \caption{The ($\ell$, $r_q$) parameters separated by the extremal case (dotted-line) $\delta_1=\delta_c$.}
    \label{fig:regions}
\end{figure}

When particularizing the above condition on $\delta_1$ for a supermassive black hole of mass $M \sim 4.2\times10^6M_\odot$ like SgrA* at the centre of the Milky Way, one gets the limit $N_q = N_c$ for an extremely low value of $r_q=3.2315\times10^{-35}$ m, when compared to scales of the astrophysical objects that we are dealing in this work. Hence, we will discard the cases $N_q \leq N_c$. In Figure \ref{fig:regions}, we show where the separation between the cases $\delta_1>\delta_c$ and $\delta_1<\delta_c$ lies on the ($\ell$, $r_q$) plane.

\section{Numerical integration of the geodesic equations}
\label{sec:numerical}

Our goal is to study whether the Galactic Center star S2 can impose constraints on the space-time metric in Eq. \eqref{eq:metric}. The dynamics of massive test particles undergoing free-fall is described by the geodesic equations found upon derivation of Eq. \eqref{eq:metric}. In particular, the world-line components $\{t(\tau),\,x(\tau),\,\theta(\tau),\,\phi(\tau)\}$ of a massive test particle trajectory in space-time satisfy the system of second-order ordinary differential equations given by
\begin{align}
    \ddot{t} = & - \frac{\dot{t} \dot{x}  A'{\left(x \right)}}{A{\left(x \right)}},\label{eq:geodesic_t}\\
    \nonumber\\
    \ddot{x} = & \dot{\phi}^{2} B{\left(x \right)} r{\left(x \right)} \sin^{2}{\left(\theta \right)}  r'{\left(x \right)} - \frac{\dot{t}^{2} B{\left(x \right)}  A'{\left(x \right)}}{2} +\nonumber \\
    &+ \dot{\theta}^{2} B{\left(x \right)} r{\left(x \right)}  r'{\left(x \right)} + \frac{\dot{x}^{2}  B'{\left(x \right)}}{2 B{\left(x \right)}},\label{eq:geodesic_x}\\
    \nonumber\\
    \ddot{\theta} = & \dot{\phi}^{2} \sin{\left(\theta \right)} \cos{\left(\theta \right)} - \frac{2 \dot{\theta} \dot{x}  r'{\left(x \right)}}{r{\left(x \right)}}
    ,\label{eq:geodesic_theta}\\
    \nonumber\\
    \ddot{\phi} &= - \frac{2 \dot{\phi} \dot{\theta} \cos{\left(\theta \right)}}{\sin{\left(\theta \right)}} - \frac{2 \dot{\phi} \dot{x}  r'{\left(x \right)}}{r{\left(x \right)}},\label{eq:geodesic_phi}
\end{align}
where dots represent derivatives with respect to the proper time $\tau$ of the particle,  and primes represent derivatives with respect to the space-time radial coordinate $x$. In particular, by indicating with $\{\dot{t}(\tau),\,\dot{x}(\tau),\,\dot{\theta}(\tau),\,\dot{\phi}(\tau)\}$ the components of the particle 4-velocity (\emph{i.e.} the tangent 4-vector to the geodesic), the normalization condition $g_{\mu\nu}\dot{x}^\mu\dot{x}^{\nu}= -1$ holds for time-like particles. Since the norm of the 4-velocity is unaltered by the parallel transport on the geodesic itself, we can assign this normalization condition at a given initial $\tau_0$, and we are assured that it will hold for the entire evolution. This practically helps in assigning initial data for the geodesic, since we can assign an initial position in space-time $\{t(\tau_0),\,x(\tau_0),\,\theta(\tau_0),\,\phi(\tau_0)\}$ (a further simplification is given by the fact that we can choose $t(\tau_0)$ arbitrarily owing to the stationarity of the metric considered) and the initial value for three out of the four components of the initial 4-velocity and automatically deriving the fourth one from the normalization condition (in our case we express $\dot{t}(0)$ as a function of $\dot{x}(0)$, $\dot{\theta}(0)$ and $\dot{\phi}(0)$). The space-time trajectory is thus characterized by 6 degrees of freedom representing the spatial position and velocity at the initial time. In celestial mechanics, it is a common choice to recast these six initial data into a set of geometrical and physical quantities, usually referred to as Keplerian orbital elements. Namely: the time of passage at pericentre $T_p$, the semi-major axis of the orbit $a$, its eccentricity $e$, the orbital inclination $i$, the angle of the line of nodes $\Omega$ and the argument of the pericentre $\omega$. The first three parameters fix the in-orbital plane motion of the test particle, while the three angular parameters define its orientation with respect to the sky-plane of a distant observer. 

A further simplification, arising from the spherical symmetry of the problem at hand, is given by the fact that, as it results from Eq. \eqref{eq:geodesic_theta}, a geodesic with $\theta = \pi/2$ and $\dot{\theta} = 0$ will have $\ddot{\theta} = 0$, identically. This means that geodesics are planar, and we can integrate the geodesic on the equatorial plane and only later perform a rotation in the distant observer reference frame. Importantly, in classical Newtonian celestial mechanics, the Keplerian orbital elements uniquely identify a closed elliptical orbit that is periodically travelled by the freely falling particle (assuming that no external perturbation is present so that its dynamics is totally regulated by the Newtonian central potential $\sim 1/r^2$ of the massive source of the gravitational field). In GR (and extensions) this is no longer valid, as the effective gravitational potential felt by the orbiting body presents higher order terms (\emph{e.g.} terms of order $\mathcal{O}(1/r^3)$ or higher) that make the massive particle follow a preceding orbital path that has the peculiar shape of a rosette \cite{Poisson2014}. This leads to a dynamical evolution of the Keplerian orbital elements over time that is encoded in the geodesic equations Eqs. \eqref{eq:geodesic_t}-\eqref{eq:geodesic_phi} that can be computed perturbatively \cite{Will1993}. Hence, for all practical purposes, the Keplerian elements at a given time, \emph{e.g.} at the initial time, identify the osculating conic curve to the actual GR-trajectory of the particle. 

In this work, we compute numerically the fully relativistic sky-projected mock trajectory for S2 by directly integrating Eqs. \eqref{eq:geodesic_t}-\eqref{eq:geodesic_phi} given a set of orbital elements. For more details, we refer to past works on the subject \cite{DeMartino2021, DellaMonica2022a, DellaMonica2022b, DellaMonica2022c, DellaMonica2023a}, while here only the main steps of our integration procedure are reported for the sake of completeness. Our orbital model is based on the following orbital parameters:
\begin{equation}
    \bm{\theta}=\left(
        \begin{array}{l}
        M,\,r_q,\,\ell,\, T_p,\, a,\, e,\, i,\, \Omega,\, \omega,\, \\ \alpha_0,\, \delta_0,\, D,\, v_\alpha^0,\, v_\delta^0,\, v_{\rm LOS}^0
        \end{array}
    \right).
    \label{eq:parameters}
\end{equation}
The first three, namely, the mass $M$ of the central object and the scale lengths $r_q$ and $\ell$, uniquely fix the space-time metric in Eq. \eqref{eq:metric} and the free parameters in the geodesic equations \eqref{eq:geodesic_t}-\eqref{eq:geodesic_phi}. Then, we have the already-mentioned Keplerian elements $(T_p, a, e, i, \Omega, \omega)$ that are converted in the initial position and velocity in the reference frame of the gravitational source. This is done by considering that $a$ and $e$ identify two radial turning points, namely the radius of pericenter $x_p = a(1-e)$ and of apocenter $x_a = a(1+e)$. These, in turn, correspond to the root of the effective potential felt by the freely falling test particle. So, fixing $a$  and $e$ uniquely fixes the specific energy and angular momentum of the geodesic that can be converted to an initial radial velocity $\dot{x}(\tau_0)$ and angular velocity $\dot{\phi}(\tau_0)$ on the equatorial plane. This further requires fixing the position in space from which the integration is started, which is done by fixing the initial time from the last pericenter passage $T_p$. In our case, without loss of generality, the initial time is taken to be the last apocentre passage, $t(\tau_0) = T_a = T_p-T/2\sim2010.35$ (being $T$ the orbital period). This completely fixes the initial position and velocity of the particle on the black hole equatorial plane, from which the geodesic equations are integrated numerically via an adaptive-step-size 4(5) Runge-Kutta integrator, both forward and backward in time (to cover the period for which observations are available).

\begin{figure*}[!ht]
    \includegraphics[width=\textwidth]{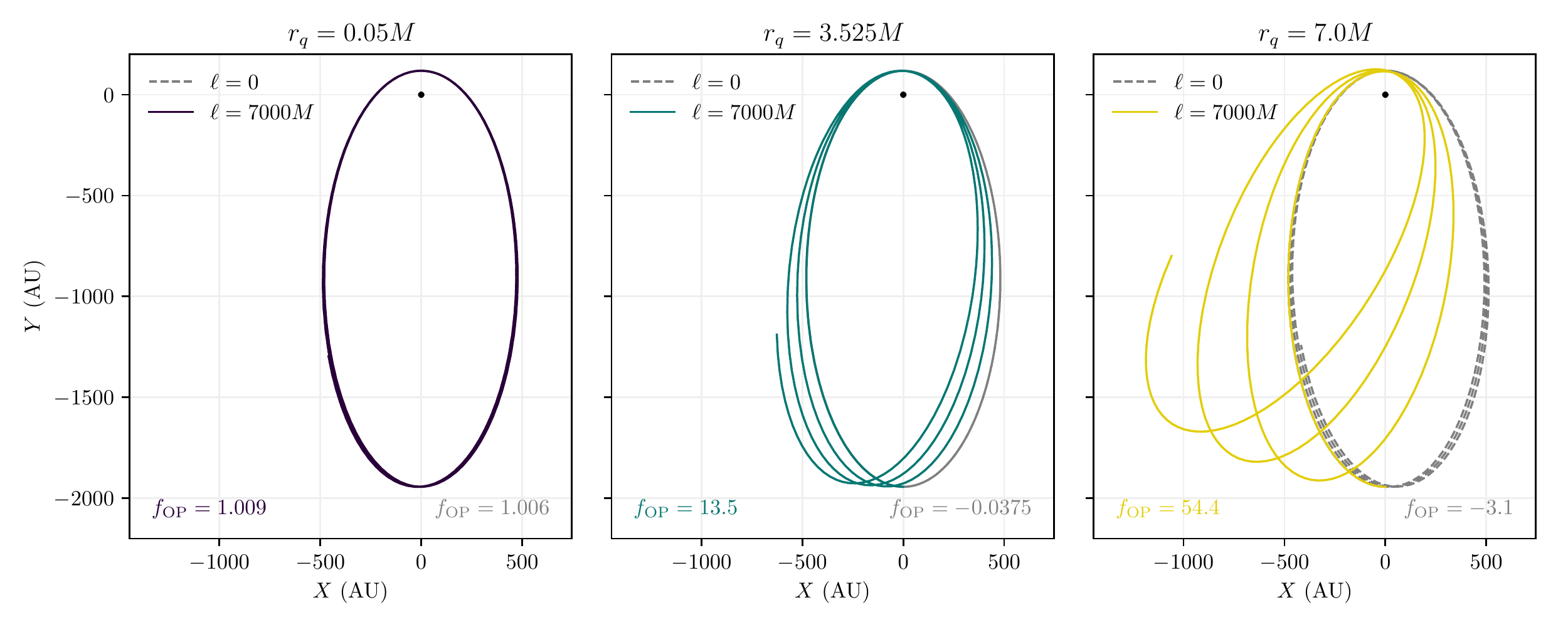}
    \caption{Numerically integrated orbits on the equatorial plane of the space-time in Eq. \eqref{eq:metric} for different values of the parameters $r_q$ and $\ell$ and for the orbital parameters of the S2 star. In particular dashed lines report the orbits for $\ell = 0$ and the value of $r_q$ reported above the plot, while solid lines show the orbit obtained for the same value of $r_q$ and a big value of $\ell = 7000M$. The text labels in the figure corners report the value of the computed rate of orbital precession $f_{\rm OP}$ with respect to the corresponding value in the Schwarzschild space-time. A negative value of $f_{\rm OP}$ corresponds to a geodesic in retrograde precession.}
    \label{fig:orbits}
\end{figure*}

The integrated orbits are subsequently converted into the physical quantities that are experimentally observed, namely: the relative right ascension ($\alpha$) and declination ($\delta$) of the star, its line of sight apparent velocity ($v_{\rm LOS}$), and the ratio of the rate of orbital precession (OP) with respect to the one predicted in GR, i.e.,
\begin{equation} \label{eq:f_OP}
f_{\rm OP} = \frac{\Delta \omega}{\Delta \omega_{GR}},
\end{equation}
where 
\begin{equation}
    \Delta\omega_{\rm GR} = \frac{6\pi GM}{c^2a(1-e^2)},
\end{equation}
During this procedure, the additional parameters considered in Eq. \eqref{eq:parameters} make their appearance in our orbital model. The transformation between the black hole reference frame to the one of the distant observer is made by a rotation by the angles $i$, $\Omega$ and $\omega$ (encoded in Thiele-Innes elements) followed by a translation by the distance $D$ of the Galatic Center from an Earth-based of observatory along the line-of-sight (LOS). In doing such transformation, the classical Rømer effect is computed and an additional time-dependent translation is considered in the direction perpendicular to the LOS (\emph{i.e.} directly on the observer's sky-plane) to account for possible offset and drift of the zero point of the astrometric of the reference frame. This systematic observational effect is encoded in the parameters $(\alpha_0,\, \delta_0,\, v_\alpha^0,\, v_\delta^0)$. The LOS velocity computation requires additional care, due to the emergence of relativistic effects. In particular, to compute a realistic value for $v_{\rm LOS}$, we first consider the kinematic velocity of the star projected along the LOS direction. Such kinematic velocity is then converted into a classical longitudinal Doppler effect which accounts for much of the observed shift. Additional redshift contributions arising from 1PN relativistic effects are taken into account, namely the special and general relativistic time dilation effects (related to the high kinematic velocity, $\sim7700$ km/s, and gravitational potential, $\Phi/c^2 = GM/rc^2\sim 3\times 10^{-4}$, of S2 at its pericenter) produce an additional redshift ($\sim 200$ km/s more in the reconstructed LOS velocity around pericenter) that has been experimentally confirmed \cite{Do:2019txf,2018A&A...618L..10G}. We estimate the additional redshift contribution from the time component of integrated four-velocity, $\dot{t} = dt/d\tau$.  Finally, a possible drift of the LOS velocity $v_{\rm LOS}^0$ due to the conversion to the local standard of rest is taken into account (here classical composition of velocities is taken into account instead of a Lorentz boost due to the intrinsic smallness of such systematic effect).
The last experimental observable, namely the rate of orbital precession $\Delta \omega$, is computed by deriving from the numerically integrated orbit the angle $\Delta \phi$ spanned by the massive test particle on the black hole equatorial plane between two consecutive radial turning points, given by definition by $2\pi+\Delta\omega$. 

\begin{figure}[!ht]
    \includegraphics[width=\columnwidth]{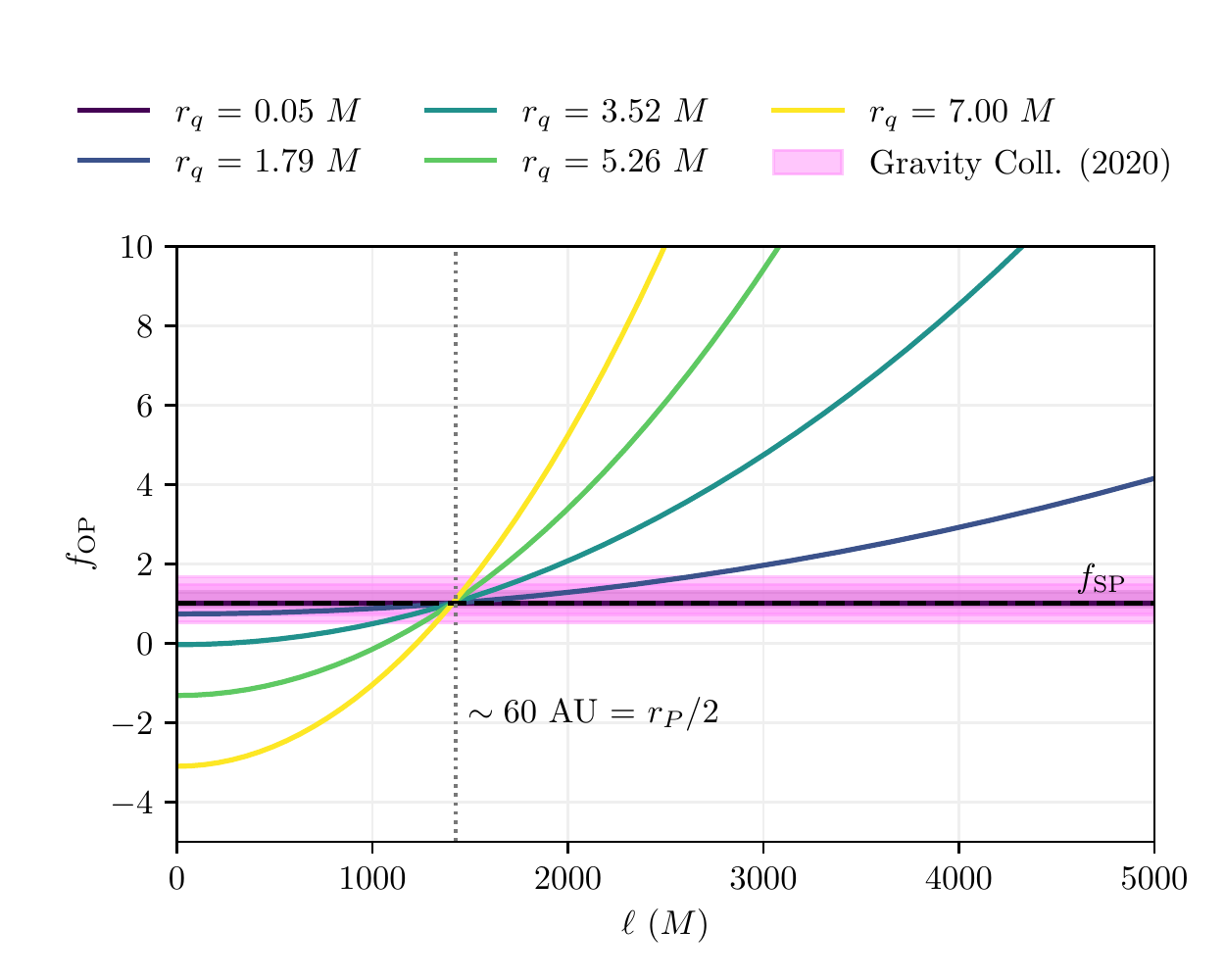}
    \caption{Ratio of the rate of orbital precession (OP) with respect to the one predicted in GR, $f_{\rm OP} = \Delta \omega/\Delta\omega_{\rm GR}$, computed for several values of the charge-related length-scale $r_q$ as a function of the curvature length-scale $\ell$.}
    \label{fig:precession}
\end{figure}

A few S2-like example trajectories from our orbital model are reported in Figure \ref{fig:orbits}, where we show the results of our numerical integration on the equatorial plane of the black hole (\emph{i.e.} before performing the sky-plane projection), for different combinations of the theory parameters. In particular, in these plots we have fixed all the orbital parameters to their known values in the literature \cite{Gillessen2017} and only changed the parameters $r_q$ and $\ell$ within their range of interest. A noticeable effect on the orbit results from the modifications of the space-time metric, which accounts for a rate of orbital precession $f_{\rm OP}$ different from unity and that, depending on the combination of parameters, can become negative (implying retrograde precession) or largely greater than one.
In more detail, in Figure \ref{fig:precession}, we report the ratio of the orbital precession in metric affine-gravity with respect to the value in GR, $f_{\rm OP}$ in Eq.(\ref{eq:f_OP}) as a function of $\ell$ for several different values of $r_q$. The horizontal pink stripe represents the confidence interval for the rate of orbital precession derived experimentally by the Gravity Collaboration after the last pericenter passage of S2 in 2018 \cite{GRAVITY:2019tuf}. As it appears, for specific combinations 
of the parameters $r_q$ and $\ell$, departures from the general relativistic precession rate may range from a factor $\sim-4$, hence implying retrograde (instead of the prograde) orbital precession, to factors greater than 10, whereas the experimentally measured value of $f_{\rm OP}$ is bound with $\sim10\%$ precision around 1. These sometimes huge departures tell us qualitatively that orbital data for S2 should be able to constrain those two extra parameters despite any degenerations with the orbital parameters. 
Moreover, for values of $\ell\sim1400 M \sim 60$ AU (coincidentally corresponding with the semi-distance of S2 from SgrA* at its pericenter passage) the orbital precession in metric-affine gravity is always coincident with the value in GR ($f_{\rm OP} =1$), regardless of the value of $r_q$, thus implying a compensation of effects in the orbital dynamics, even for underlying geometries that depart greatly from that of a Schwarzschild black hole.

\section{Data and Data Analysis}
\label{sec:data}

The S2 star orbits the radio source SgrA* in the center of the Milky Way with an orbital period of $\sim 16$ years, a semi-major axis of  $\sim$ 1000 AU, and a high eccentricity of $\sim 0.88$ \cite{DeLaurentis2023}. At periastron the distance from SgrA* reduces to about $\sim$ 120 AU, meaning that the star is barely 1400 gravitational radii away from the central compact object. Moreover, 
during the passage to the pericenter, the orbital velocity reaches $\sim 7700$ km/s (which is $\sim2.5\%$ of the speed of light in vacuum). For these reasons, the S2 pericenter passage has proven to be a crucial test bench for relativistic effects in the local universe \cite{2018A&A...618L..10G, GRAVITY:2020xcu, Do:2019txf}. Correspondingly, the tight constraints imposed by the S2 orbit on the validity of relativistic predictions have been translated into bounds on possible modifications to the underlying theory of gravity or alternatives to the standard black hole paradigm \cite{DeMartino2021, DellaMonica2022a, DellaMonica2022b, DellaMonica2023a, Cadoni2023, DeLaurentis2023}. 

We use publicly available data for the S2 star to derive constraints on the orbital model introduced in Section \ref{sec:numerical} related to the metric in Eq. \eqref{eq:metric}. In particular, while the entire set of parameters in Eq. \eqref{eq:parameters} is fitted for by our analysis, we 
focus on the parameters $r_q$ and $\ell$ that encode metric deviations from the standard Schwarzschild black hole.

Our dataset is composed of three distinct kinds of measurements:
\begin{itemize}
    \item \textbf{Astrometric positions:} right ascension ($\alpha$) and declination ($\delta$) recorded at 145 epochs between $\sim1992$ and $\sim2016$ at ESO facilities, with $\approx 400 \mu$as of average experimental uncertainty. Positions are reported relative to the ‘GC radio-to-infrared reference system’ \cite{Plewa2015}, \emph{i.e.} relative to the point on the sky where radio observations pinpoint the position of SgrA*. Uncertainty on this estimate is accounted for in the parameters $(\alpha_0,\, \delta_0,\, v_\alpha^0,\, v_\delta^0)$ introduced in Eq. \eqref{eq:parameters} to 
    consider a possible offset and drift of the zero point of the astrometric reference frame \cite{Gillessen2009a, Gillessen2017, Plewa2015}. These data are publicly available in \cite{Gillessen2017}.
    \item \textbf{Line-of-sight velocities} ($v_{\rm LOS}$):  the measured line-of-sight velocity of the S2 (assumed positive during the approaching phase and negative during the recession) over a total of 44 epochs in the same period as the astrometric positions. The correction offset of these observations to the local standard of rest is accounted for by the parameter $v_{\rm LOS}^0$ in Eq. \eqref{eq:parameters}.
    \item \textbf{Orbital precession} ($f_{\rm SP}$): the determination of the rate of orbital precession $f_{\rm OP}$ made possible by the measurement, during and after the 2018 pericenter passage of the S2 star, of astrometric positions with unprecedented accuracy by the Gravity Collaboration \cite{GRAVITY:2020gka} using the GRAVITY interferometer at VLT. Such astrometric dataset is not publicly available, so we introduce the measured rate of orbital precession\footnote{Here, we use the notation $f_{\rm SP}$ to refer to the measurement by the Gravity Collaboration of $f_{\rm OP}$ in Eq. \eqref{eq:f_OP}. The ``SP'' conventionally stands for Schwarzschild Precession, as what has been measured is the accordance of observed position for S2 to what a Schwarzschild model for the central black hole would predict.}, $f_{\rm SP}=1.10\pm 0.19$, as an additional data-point to our dataset.
    Such a measurement excludes Newtonian gravity (which would imply $f_{\rm OP}= 0$) at 5$\sigma$ and shows perfect agreement with the GR prediction for a Schwarzschild black hole. 
    In Figure \ref{fig:precession}, we have already compared our predicted rate of precession, $f_{\rm OP}$, for different combinations of the parameters  $r_q$ and $\ell$ with the Schwarzschild one, $f_{\rm SP}$, measured by the Gravity Collaboration.
\end{itemize}
    
We carry out a Markov Chain Monte Carlo (MCMC) analysis to explore the parameter space in Eq. \eqref{eq:parameters} for our orbital model. In particular, we use the \texttt{emcee} \cite{Foreman2013} Bayesian sampler, by assigning flat priors for the Keplerian orbital elements of S2 centered on the best-fit values from \cite{Gillessen2017} and spanning a range 10 times larger than the uncertainty estimated in the same analysis. For the parameters related to systematic effects on the reference frame, we assign Gaussian priors with central values and amplitudes taken from the independent analysis in \cite{Plewa2015}. Finally, for the parameters of interest of our analysis, namely $r_q$ and $\ell$, we set the uniform priors $r_q\in[0,\,7]M$ and $\ell\in[0,\,5000]M$, that have been chosen heuristically. For each extracted set of parameters we quantify the agreement between the observational data and our orbital model by the following likelihood
\begin{widetext}
    \begin{align}
        \log \mathcal{L} =& -\frac{1}{2}\biggl[\sum_i^{145}\biggl(\frac{\alpha_i(\bm{\theta})-\alpha^{{obs}}_{i}}{\sqrt{2}\sigma_{\alpha,i}}\biggr)^2+\sum_i^{145}\biggl(\frac{\delta_i(\bm{\theta})-\delta^{{obs}}_{i}}{\sqrt{2}\sigma_{\delta,i}}\biggr)^2+ \sum_i^{44}\biggl(\frac{v_{{\rm LOS}, i}(\bm{\theta})-v_{{\rm LOS}, i}^{obs}}{\sqrt{2}\sigma_{v_{\rm LOS},i}}\biggr)^2+
        \biggl(\frac{f_{\rm OP}(\bm{\theta})-f_{{\rm SP}}}{\sqrt{2}\sigma_{f_{\rm SP}}}\biggr)^2\biggr].
        \label{eq:likelihood_precession}
    \end{align}
\end{widetext}
The labels $obs$ indicate observed quantities from \cite{Gillessen2017} at the $i$-th epoch (with $i$ spanning from 1 to 145 for the astrometric positions and from 1 to 45 for the line-of-sight velocities), while unlabeled quantities indicate the prediction from our orbital model for a given set of parameters $\bm{\theta}$. The $\sigma$s at the denominators are the corresponding uncertainties. Since the measurement of the rate of orbital precession $f_{\rm SP}$ by \cite{GRAVITY:2020gka} involves the same dataset used here by \cite{Gillessen2017} (plus the more accurate observations at and after the pericenter passage in 2018), we conservatively added the $\sqrt{2}$ in the denominators of Eq. \eqref{eq:likelihood_precession}, to avoid double counting data \cite{DeMartino2021}.

\section{Results}
\label{sec:results}
\begin{figure*}[t]
    \centering
    \includegraphics[width=\textwidth]{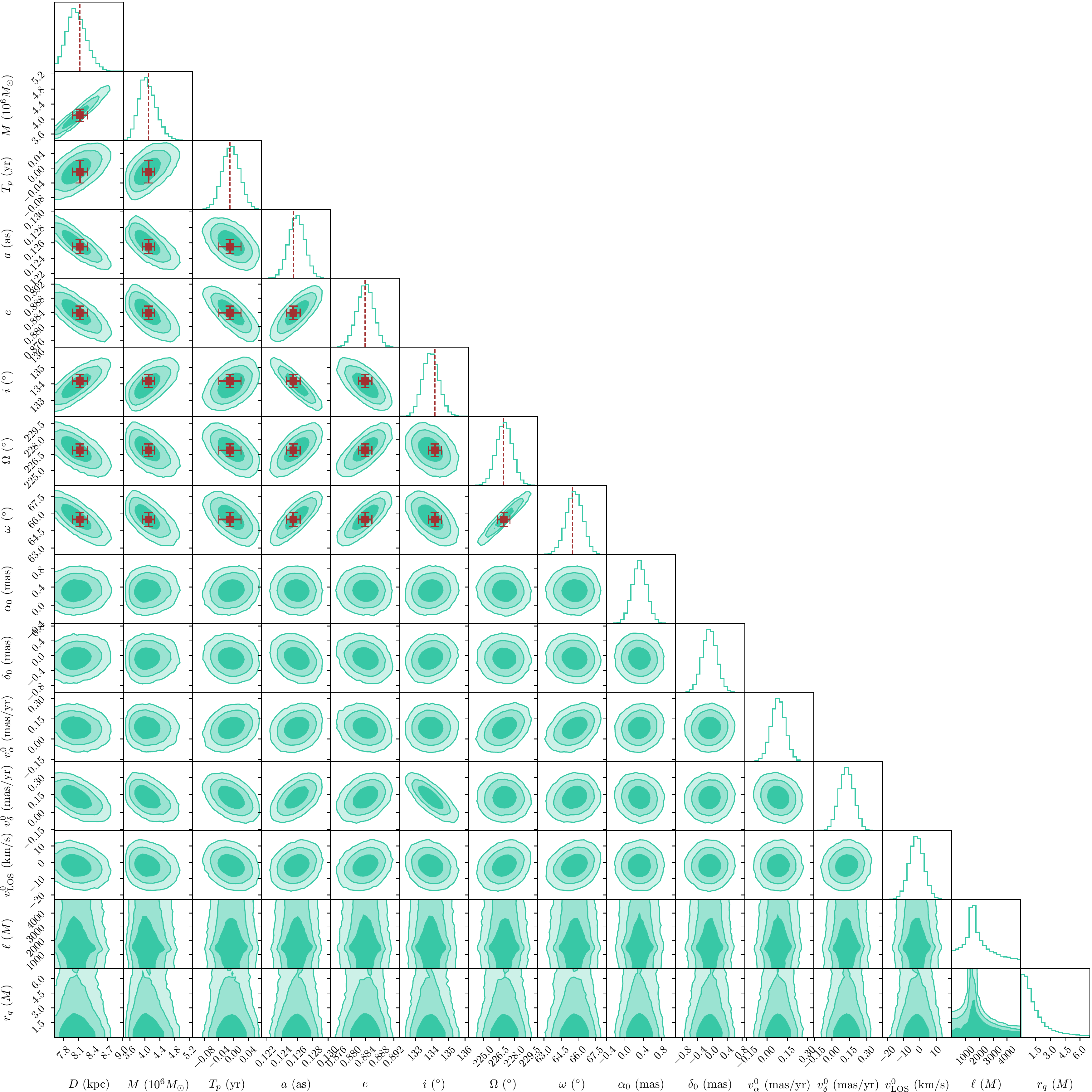}
    \caption{Full 15-dimensional posterior distribution from our MCMC analysis. On the diagonal, we report marginalized 1d posterior distributions for each parameter in our orbital model for S2 in metric-affine gravity, while for each pair of parameters, we report 2d contour plots embracing (from darker to lighter) the 68\%, 95\% and 99.7\% of the posterior samples in the off-diagonal plots. Red marks report the best fitting value for the orbital parameters of S2 from the Newtonian analysis in \cite{Gillessen2017} with the corresponding 1$\sigma$ uncertainties.}
    \label{fig:corner_plot}
\end{figure*}

\begin{table}[t]
    \setlength{\tabcolsep}{20pt}
    \renewcommand{\arraystretch}{1.5}
    \begin{tabular}{|lc|}
        \hline
        Parameter (units) & Best-fit \\ \hline
        $M$ ($10^6M_\odot$) & $4.08\pm0.23$ \\
        $D$ (kpc) & $8.07\pm0.2$ \\
        $T_p-2018.37$ (yr) & $-0.007\pm0.023$ \\
        $a$ (as) & $0.126\pm0.001$ \\
        $e$ & $0.884\pm0.0023$ \\
        $i$ ($^\circ$) & $133.93\pm0.48$ \\
        $\Omega$ ($^\circ$) & $227.01\pm0.77$ \\
        $\omega$ ($^\circ$) & $65.67_{-0.71}^{+0.72}$ \\
        $\alpha_0$ (mas) & $0.32\pm0.16$ \\
        $\delta_0$ (mas) & $-0.07\pm0.19$ \\
        $v_{\alpha}^0$ (mas/yr) & $0.084\pm0.052$ \\
        $v_{\delta}^0$ (mas/yr) & $0.126\pm0.063$ \\
        $v_{\rm LOS}^0$ (km/s) & $-2.0\pm4.4$ \\ \hline
    \end{tabular}
    \caption{Resulting $68\%$ confidence intervals for all the orbital parameters from our MCMC analysis. All orbital parameters are bound and consistent within $1\sigma$ with their best fitting values from other analyses in the literature \cite{Gillessen2017}.}
    \label{tab:posteriors}
\end{table}

\begin{figure}[!ht]
    \includegraphics[width=\columnwidth]{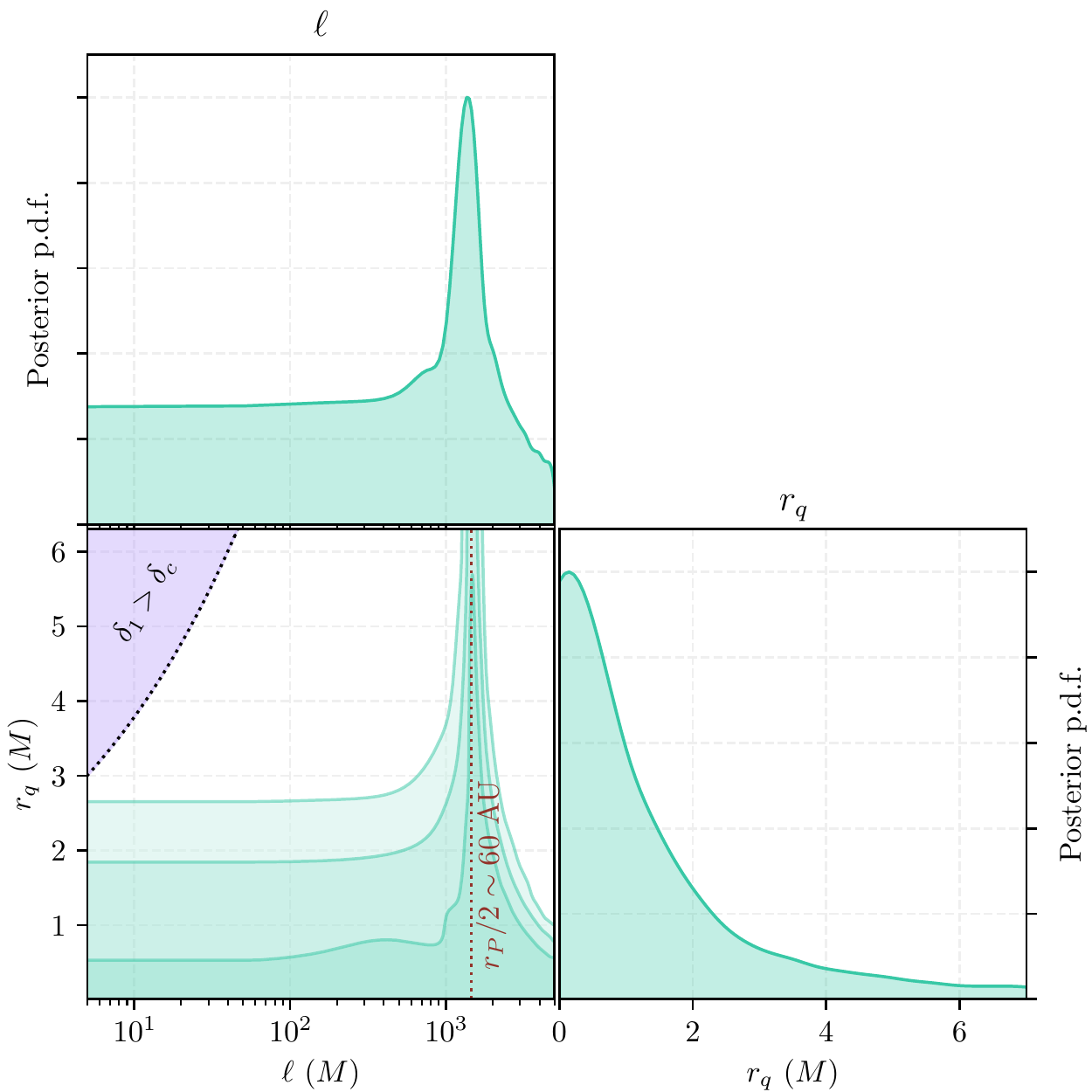}
    \caption{Posterior probability distribution for the parameters of interest of our analysis, namely $r_q$ and $\ell$. In particular, the top and right panels show the marginalized posterior distributions for the two parameters, while the bottom-left panel shows the 2d-slice of the full posterior distribution corresponding to the $r_q$-$\ell$ plane. The 1$\sigma$, 2$\sigma$ and 3$\sigma$ regions are highlighted with increasingly darker shades. Moreover, the vertical dashed red line corresponds to a value of $\ell \sim r_p/2$. The shaded violet region corresponds to the class of models with $\delta_1 > \delta_c$. Remarkably, our analysis bounds the parameter space in the region $\delta_1 < \delta_c$, showing that the orbit of S2 is sensible to such strong modifications of the space-time metric.}
    \label{fig:posteriors}
\end{figure}

In this section, we present the results of our Bayesian analysis. In Table \ref{tab:posteriors}, we report the median-centered 68\% confidence intervals on the orbital and reference frame parameters of our orbital model retrieved by the sampled posterior distribution. 
The effectiveness of our analysis can be easily noted in Figure \ref{fig:corner_plot} where we show, in a corner plot, all the marginalized posterior distributions of our parameter space along the diagonal, and the 2D contours for all possible couples of parameters depicting the 68\%, 95\%  and 99.7\% credible regions. All the parameters are bound and the confidence intervals are compatible with those of previous analyses \cite{Gillessen2017}, as shown by the red square points, with their corresponding error bars, indicating their best-fitting values. This confirms the validity of our analysis in recovering the appropriate set of orbital parameters for S2.

The confidence regions for our parameters of interest for the metric-affine space-time solution in Eq. \eqref{eq:metric} are also shown in Figure \ref{fig:posteriors}. Here, in the contour plot, we show the interesting allowed regions (at 1, 2 and 3$\sigma$ confidence) for the parameters $r_q$ and $\ell$ that we have derived as a result of our analysis, while the marginalized posterior distributions for the two parameters are reported on the diagonal of the corner plot. Our orbital model for S2 tends to prefer small values of $r_q$, generally below $2.7M$, for values of $\ell$ below the semi-radius of the pericenter passage of S2, at around 60 AU (reported as a red dotted vertical line in our contour). For $\ell=60$ AU, the orbit of S2 appears to be completely insensible of the particular value of $r_q$, as already highlighted in our analysis of the orbital precession as a function of $\ell$ and $r_q$ (see Figure \ref{fig:precession}). For $\ell>60$ AU the deviations to the orbital dynamics of S2 are strong even for very small values of $r_q$, which brings the credible interval for this parameter to narrow down, with $r_q < M$ at $\lambda \sim 5000M$. Quite remarkably, our analysis constrains $r_q$ and $\ell$ within the $\delta_1<\delta_c$ region, implying that the orbit for S2 favours a central object that resembles a Schwarzschild black hole when the gravitational field is described by the space-time metric in Eq. \eqref{eq:metric}. This excludes, in the range of the priors that we have considered, the severe modifications (in terms of their causal structures) to the underlying geometries predicted for the cases $\delta_1\geq\delta_c$ (highlighted in violet in the contour plot).

\section{Discussion and remarks}\label{sec:conclusions}

The remarkable precision of astronomical observations of the S-stars at the Galactic Center has been a crucial piece of evidence towards the identification of SgrA* as a supermassive black hole residing at the heart of the Milky Way. Such measurements have led to the first detection of orbital relativistic effects, and thus an indirect demonstration of the general relativistic geodesic paradigm for the motion of freely falling test particles, outside of our Solar System \cite{GravityCollaboration2018a, Do:2019txf}. The ability to perform metric tests of gravity using these observations has motivated numerous studies that have leveraged such opportunity to translate the narrow margins of deviations from General Relativity, naturally encoded in the observational uncertainties, into exclusion regions for several modified theories of gravity \cite{Borka2012, Borka2013, DAddio2021, DeMartino2021, DellaMonica2022a, DellaMonica2023a, DellaMonica2023b}, black hole mimickers \cite{DellaMonica2022b, Cadoni2023} and dark matter models \cite{DellaMonica2022c, DellaMonica2023c, GRAVITY:2023cjt}. 

In this work we have used the publicly available data for the star S2, the brightest in the cluster, to derive the first constraints based on such a data on quadratic modifications to the General Relativistic action in the metric-affine formalism, resulting in the space-time geometry reported in Eq. \eqref{eq:metric}. We have developed an orbital model, based on the numerical integration of the geodesic equations in Eqs. \eqref{eq:geodesic_t}-\eqref{eq:geodesic_phi} that takes into account all relativistic and systematic effects necessary to model the S2 orbital data accurately \cite{Gillessen2017, GRAVITY:2020hwn}. Our analysis, which recovers values for the orbital parameters within 1$\sigma$ (as shown in Figure \ref{fig:corner_plot} and Table \ref{tab:posteriors}) that agree with past studies, resulted in interesting constraints on the extra length-scales by which this family of solutions is characterized, namely, a charge-related length $r_q$ and a parameter modulating the higher-curvature terms in the Lagrangian $\ell$, whose results are reported in Figure \ref{fig:posteriors}. The parameter $r_q$ is generally bound below $2.7M$, except when $\ell \sim 60$ AU (coinciding with the semi-distance of S2 from SgrA* at pericenter passage) in which case the orbit of S2 appears to be insensible of the particular value of $r_q$ (as also qualitatively seen analyzing the orbital precession for S2 as a function of $\ell$ and $r_q$ in Figure \ref{fig:precession}). This analysis, which provides the first constraints for the metric element in Eq. \eqref{eq:metric} at the scales of the Galactic Center, provides a remarkable complementary result to past tests on the same theory \cite{Chen2019, Rubiera2020, Bahamonde2021, Olmo2022} and on the subject of regular black holes in general \cite{Bambi2023, Lan2023}. 

The fact that our results single out the Schwarzschild-like subfamily of metric-affine black holes (i.e. those with $\delta_1<\delta_c$) is consistent with the results from the analysis of other messengers such as gravitational waves and shadow images out of accretion disks, all of them pinpointing that hard modifications of the black hole paradigm of GR may be strongly disfavoured by data. At the same time, it also reinforces the need to cross-test several such messengers to determine the viability of any alternative to the canonical black hole paradigm of GR, like the metric-affine one considered in this work.

\section*{Acknowledgements}
IDM and RDM acknowledge support from the grant PID2021-122938NB-I00, and DRG acknowledges support from the grant PID2022-138607NB-I00, both funded by MCIN/AEI/10.13039/501100011033 and by ``ERDF A way of making Europe''.  RDM also acknowledges support from Consejeria de Educación de la Junta de Castilla y León and the European Social Fund.



\bibliography{refs} 

\begin{thebibliography}{64}%
\makeatletter
\providecommand \@ifxundefined [1]{%
 \@ifx{#1\undefined}
}%
\providecommand \@ifnum [1]{%
 \ifnum #1\expandafter \@firstoftwo
 \else \expandafter \@secondoftwo
 \fi
}%
\providecommand \@ifx [1]{%
 \ifx #1\expandafter \@firstoftwo
 \else \expandafter \@secondoftwo
 \fi
}%
\providecommand \natexlab [1]{#1}%
\providecommand \enquote  [1]{``#1''}%
\providecommand \bibnamefont  [1]{#1}%
\providecommand \bibfnamefont [1]{#1}%
\providecommand \citenamefont [1]{#1}%
\providecommand \href@noop [0]{\@secondoftwo}%
\providecommand \href [0]{\begingroup \@sanitize@url \@href}%
\providecommand \@href[1]{\@@startlink{#1}\@@href}%
\providecommand \@@href[1]{\endgroup#1\@@endlink}%
\providecommand \@sanitize@url [0]{\catcode `\\12\catcode `\$12\catcode
  `\&12\catcode `\#12\catcode `\^12\catcode `\_12\catcode `\%12\relax}%
\providecommand \@@startlink[1]{}%
\providecommand \@@endlink[0]{}%
\providecommand \url  [0]{\begingroup\@sanitize@url \@url }%
\providecommand \@url [1]{\endgroup\@href {#1}{\urlprefix }}%
\providecommand \urlprefix  [0]{URL }%
\providecommand \Eprint [0]{\href }%
\providecommand \doibase [0]{http://dx.doi.org/}%
\providecommand \selectlanguage [0]{\@gobble}%
\providecommand \bibinfo  [0]{\@secondoftwo}%
\providecommand \bibfield  [0]{\@secondoftwo}%
\providecommand \translation [1]{[#1]}%
\providecommand \BibitemOpen [0]{}%
\providecommand \bibitemStop [0]{}%
\providecommand \bibitemNoStop [0]{.\EOS\space}%
\providecommand \EOS [0]{\spacefactor3000\relax}%
\providecommand \BibitemShut  [1]{\csname bibitem#1\endcsname}%
\let\auto@bib@innerbib\@empty
\bibitem [{\citenamefont {Abbott}\ \emph {et~al.}(2016)\citenamefont {Abbott}
  \emph {et~al.}}]{LIGOScientific:2016aoc}%
  \BibitemOpen
  \bibfield  {author} {\bibinfo {author} {\bibfnamefont {B.~P.}\ \bibnamefont
  {Abbott}} \emph {et~al.} (\bibinfo {collaboration} {LIGO Scientific,
  Virgo}),\ }\href {\doibase 10.1103/PhysRevLett.116.061102} {\bibfield
  {journal} {\bibinfo  {journal} {Phys. Rev. Lett.}\ }\textbf {\bibinfo
  {volume} {116}},\ \bibinfo {pages} {061102} (\bibinfo {year} {2016})},\
  \Eprint {http://arxiv.org/abs/1602.03837} {arXiv:1602.03837 [gr-qc]}
  \BibitemShut {NoStop}%
\bibitem [{\citenamefont {Akiyama}\ \emph {et~al.}(2019)\citenamefont {Akiyama}
  \emph {et~al.}}]{EventHorizonTelescope:2019dse}%
  \BibitemOpen
  \bibfield  {author} {\bibinfo {author} {\bibfnamefont {K.}~\bibnamefont
  {Akiyama}} \emph {et~al.} (\bibinfo {collaboration} {Event Horizon
  Telescope}),\ }\href {\doibase 10.3847/2041-8213/ab0ec7} {\bibfield
  {journal} {\bibinfo  {journal} {Astrophys. J. Lett.}\ }\textbf {\bibinfo
  {volume} {875}},\ \bibinfo {pages} {L1} (\bibinfo {year} {2019})},\ \Eprint
  {http://arxiv.org/abs/1906.11238} {arXiv:1906.11238 [astro-ph.GA]}
  \BibitemShut {NoStop}%
\bibitem [{\citenamefont {Akiyama}\ \emph {et~al.}(2022)\citenamefont {Akiyama}
  \emph {et~al.}}]{EventHorizonTelescope:2022wkp}%
  \BibitemOpen
  \bibfield  {author} {\bibinfo {author} {\bibfnamefont {K.}~\bibnamefont
  {Akiyama}} \emph {et~al.} (\bibinfo {collaboration} {Event Horizon
  Telescope}),\ }\href {\doibase 10.3847/2041-8213/ac6674} {\bibfield
  {journal} {\bibinfo  {journal} {Astrophys. J. Lett.}\ }\textbf {\bibinfo
  {volume} {930}},\ \bibinfo {pages} {L12} (\bibinfo {year}
  {2022})}\BibitemShut {NoStop}%
\bibitem [{\citenamefont {de~Laurentis}\ \emph {et~al.}(2023)\citenamefont
  {de~Laurentis}, \citenamefont {De~Martino},\ and\ \citenamefont
  {Della~Monica}}]{deLaurentis:2022oqa}%
  \BibitemOpen
  \bibfield  {author} {\bibinfo {author} {\bibfnamefont {M.}~\bibnamefont
  {de~Laurentis}}, \bibinfo {author} {\bibfnamefont {I.}~\bibnamefont
  {De~Martino}}, \ and\ \bibinfo {author} {\bibfnamefont {R.}~\bibnamefont
  {Della~Monica}},\ }\href {\doibase 10.1088/1361-6633/ace91b} {\bibfield
  {journal} {\bibinfo  {journal} {Rept. Prog. Phys.}\ }\textbf {\bibinfo
  {volume} {86}},\ \bibinfo {pages} {104901} (\bibinfo {year} {2023})},\
  \Eprint {http://arxiv.org/abs/2211.07008} {arXiv:2211.07008 [astro-ph.GA]}
  \BibitemShut {NoStop}%
\bibitem [{\citenamefont {Eckart}\ and\ \citenamefont
  {Genzel}(1996)}]{Eckart:1996zz}%
  \BibitemOpen
  \bibfield  {author} {\bibinfo {author} {\bibfnamefont {A.}~\bibnamefont
  {Eckart}}\ and\ \bibinfo {author} {\bibfnamefont {R.}~\bibnamefont
  {Genzel}},\ }\href {\doibase 10.1038/383415a0} {\bibfield  {journal}
  {\bibinfo  {journal} {Nature}\ }\textbf {\bibinfo {volume} {383}},\ \bibinfo
  {pages} {415} (\bibinfo {year} {1996})}\BibitemShut {NoStop}%
\bibitem [{\citenamefont {Ghez}\ \emph {et~al.}(2003)\citenamefont {Ghez} \emph
  {et~al.}}]{Ghez:2003rt}%
  \BibitemOpen
  \bibfield  {author} {\bibinfo {author} {\bibfnamefont {A.~M.}\ \bibnamefont
  {Ghez}} \emph {et~al.},\ }\href {\doibase 10.1086/374804} {\bibfield
  {journal} {\bibinfo  {journal} {Astrophys. J. Lett.}\ }\textbf {\bibinfo
  {volume} {586}},\ \bibinfo {pages} {L127} (\bibinfo {year} {2003})},\ \Eprint
  {http://arxiv.org/abs/astro-ph/0302299} {arXiv:astro-ph/0302299} \BibitemShut
  {NoStop}%
\bibitem [{\citenamefont {Abuter}\ \emph {et~al.}(2018)\citenamefont {Abuter}
  \emph {et~al.}}]{GRAVITY:2018ofz}%
  \BibitemOpen
  \bibfield  {author} {\bibinfo {author} {\bibfnamefont {R.}~\bibnamefont
  {Abuter}} \emph {et~al.} (\bibinfo {collaboration} {GRAVITY}),\ }\href
  {\doibase 10.1051/0004-6361/201833718} {\bibfield  {journal} {\bibinfo
  {journal} {Astron. Astrophys.}\ }\textbf {\bibinfo {volume} {615}},\ \bibinfo
  {pages} {L15} (\bibinfo {year} {2018})},\ \Eprint
  {http://arxiv.org/abs/1807.09409} {arXiv:1807.09409 [astro-ph.GA]}
  \BibitemShut {NoStop}%
\bibitem [{\citenamefont {Do}\ \emph {et~al.}(2019)\citenamefont {Do} \emph
  {et~al.}}]{Do:2019txf}%
  \BibitemOpen
  \bibfield  {author} {\bibinfo {author} {\bibfnamefont {T.}~\bibnamefont {Do}}
  \emph {et~al.},\ }\href {\doibase 10.1126/science.aav8137} {\bibfield
  {journal} {\bibinfo  {journal} {Science}\ }\textbf {\bibinfo {volume}
  {365}},\ \bibinfo {pages} {664} (\bibinfo {year} {2019})},\ \Eprint
  {http://arxiv.org/abs/1907.10731} {arXiv:1907.10731 [astro-ph.GA]}
  \BibitemShut {NoStop}%
\bibitem [{\citenamefont {Abuter}\ \emph
  {et~al.}(2020{\natexlab{a}})\citenamefont {Abuter} \emph
  {et~al.}}]{GRAVITY:2020gka}%
  \BibitemOpen
  \bibfield  {author} {\bibinfo {author} {\bibfnamefont {R.}~\bibnamefont
  {Abuter}} \emph {et~al.} (\bibinfo {collaboration} {GRAVITY}),\ }\href
  {\doibase 10.1051/0004-6361/202037813} {\bibfield  {journal} {\bibinfo
  {journal} {Astron. Astrophys.}\ }\textbf {\bibinfo {volume} {636}},\ \bibinfo
  {pages} {L5} (\bibinfo {year} {2020}{\natexlab{a}})},\ \Eprint
  {http://arxiv.org/abs/2004.07187} {arXiv:2004.07187 [astro-ph.GA]}
  \BibitemShut {NoStop}%
\bibitem [{\citenamefont {Senovilla}\ and\ \citenamefont
  {Garfinkle}(2015)}]{Senovilla:2014gza}%
  \BibitemOpen
  \bibfield  {author} {\bibinfo {author} {\bibfnamefont {J.~M.~M.}\
  \bibnamefont {Senovilla}}\ and\ \bibinfo {author} {\bibfnamefont
  {D.}~\bibnamefont {Garfinkle}},\ }\href {\doibase
  10.1088/0264-9381/32/12/124008} {\bibfield  {journal} {\bibinfo  {journal}
  {Class. Quant. Grav.}\ }\textbf {\bibinfo {volume} {32}},\ \bibinfo {pages}
  {124008} (\bibinfo {year} {2015})},\ \Eprint {http://arxiv.org/abs/1410.5226}
  {arXiv:1410.5226 [gr-qc]} \BibitemShut {NoStop}%
\bibitem [{\citenamefont {Penrose}(1965)}]{Penrose:1964wq}%
  \BibitemOpen
  \bibfield  {author} {\bibinfo {author} {\bibfnamefont {R.}~\bibnamefont
  {Penrose}},\ }\href {\doibase 10.1103/PhysRevLett.14.57} {\bibfield
  {journal} {\bibinfo  {journal} {Phys. Rev. Lett.}\ }\textbf {\bibinfo
  {volume} {14}},\ \bibinfo {pages} {57} (\bibinfo {year} {1965})}\BibitemShut
  {NoStop}%
\bibitem [{\citenamefont {{Hawking}}(1966)}]{Hawking1966}%
  \BibitemOpen
  \bibfield  {author} {\bibinfo {author} {\bibfnamefont {S.~W.}\ \bibnamefont
  {{Hawking}}},\ }\href {\doibase 10.1103/PhysRevLett.17.444} {\bibfield
  {journal} {\bibinfo  {journal} {\prl}\ }\textbf {\bibinfo {volume} {17}},\
  \bibinfo {pages} {444} (\bibinfo {year} {1966})}\BibitemShut {NoStop}%
\bibitem [{\citenamefont {{Tipler}}(1977)}]{Tipler1977}%
  \BibitemOpen
  \bibfield  {author} {\bibinfo {author} {\bibfnamefont {F.~J.}\ \bibnamefont
  {{Tipler}}},\ }\href {\doibase 10.1103/PhysRevD.15.942} {\bibfield  {journal}
  {\bibinfo  {journal} {\prd}\ }\textbf {\bibinfo {volume} {15}},\ \bibinfo
  {pages} {942} (\bibinfo {year} {1977})}\BibitemShut {NoStop}%
\bibitem [{\citenamefont {{Ellis}}\ and\ \citenamefont
  {{Schmidt}}(1977)}]{Ellis1977}%
  \BibitemOpen
  \bibfield  {author} {\bibinfo {author} {\bibfnamefont {G.~F.~R.}\
  \bibnamefont {{Ellis}}}\ and\ \bibinfo {author} {\bibfnamefont {B.~G.}\
  \bibnamefont {{Schmidt}}},\ }\href {\doibase 10.1007/BF00759240} {\bibfield
  {journal} {\bibinfo  {journal} {General Relativity and Gravitation}\ }\textbf
  {\bibinfo {volume} {8}},\ \bibinfo {pages} {915} (\bibinfo {year}
  {1977})}\BibitemShut {NoStop}%
\bibitem [{\citenamefont {Bardeen}(1968)}]{Bardeen1968}%
  \BibitemOpen
  \bibfield  {author} {\bibinfo {author} {\bibfnamefont {J.~M.}\ \bibnamefont
  {Bardeen}},\ }\href@noop {} {\enquote {\bibinfo {title} {Proceedings of the
  international conference gr5},}\ } (\bibinfo {year} {1968})\BibitemShut
  {NoStop}%
\bibitem [{\citenamefont {{Ay{\'o}n-Beato}}\ and\ \citenamefont
  {{Garc{\'\i}a}}(1998)}]{AyonBeato1998}%
  \BibitemOpen
  \bibfield  {author} {\bibinfo {author} {\bibfnamefont {E.}~\bibnamefont
  {{Ay{\'o}n-Beato}}}\ and\ \bibinfo {author} {\bibfnamefont {A.}~\bibnamefont
  {{Garc{\'\i}a}}},\ }\href {\doibase 10.1103/PhysRevLett.80.5056} {\bibfield
  {journal} {\bibinfo  {journal} {\prl}\ }\textbf {\bibinfo {volume} {80}},\
  \bibinfo {pages} {5056} (\bibinfo {year} {1998})},\ \Eprint
  {http://arxiv.org/abs/gr-qc/9911046} {arXiv:gr-qc/9911046 [gr-qc]}
  \BibitemShut {NoStop}%
\bibitem [{\citenamefont {{Bronnikov}}\ and\ \citenamefont
  {{Fabris}}(2006)}]{Bronnikov2006}%
  \BibitemOpen
  \bibfield  {author} {\bibinfo {author} {\bibfnamefont {K.~A.}\ \bibnamefont
  {{Bronnikov}}}\ and\ \bibinfo {author} {\bibfnamefont {J.~C.}\ \bibnamefont
  {{Fabris}}},\ }\href {\doibase 10.1103/PhysRevLett.96.251101} {\bibfield
  {journal} {\bibinfo  {journal} {\prl}\ }\textbf {\bibinfo {volume} {96}},\
  \bibinfo {eid} {251101} (\bibinfo {year} {2006})},\ \Eprint
  {http://arxiv.org/abs/gr-qc/0511109} {arXiv:gr-qc/0511109 [gr-qc]}
  \BibitemShut {NoStop}%
\bibitem [{\citenamefont {{Bambi}}(2023)}]{Bambi2023}%
  \BibitemOpen
  \bibfield  {author} {\bibinfo {author} {\bibfnamefont {C.}~\bibnamefont
  {{Bambi}}},\ }\href {\doibase 10.48550/arXiv.2307.13249} {\bibfield
  {journal} {\bibinfo  {journal} {arXiv e-prints}\ ,\ \bibinfo {eid}
  {arXiv:2307.13249}} (\bibinfo {year} {2023})},\ \Eprint
  {http://arxiv.org/abs/2307.13249} {arXiv:2307.13249 [gr-qc]} \BibitemShut
  {NoStop}%
\bibitem [{\citenamefont {Sotiriou}\ and\ \citenamefont
  {Faraoni}(2010)}]{Sotiriou:2008rp}%
  \BibitemOpen
  \bibfield  {author} {\bibinfo {author} {\bibfnamefont {T.~P.}\ \bibnamefont
  {Sotiriou}}\ and\ \bibinfo {author} {\bibfnamefont {V.}~\bibnamefont
  {Faraoni}},\ }\href {\doibase 10.1103/RevModPhys.82.451} {\bibfield
  {journal} {\bibinfo  {journal} {Rev. Mod. Phys.}\ }\textbf {\bibinfo {volume}
  {82}},\ \bibinfo {pages} {451} (\bibinfo {year} {2010})},\ \Eprint
  {http://arxiv.org/abs/0805.1726} {arXiv:0805.1726 [gr-qc]} \BibitemShut
  {NoStop}%
\bibitem [{\citenamefont {Clifton}\ \emph {et~al.}(2012)\citenamefont
  {Clifton}, \citenamefont {Ferreira}, \citenamefont {Padilla},\ and\
  \citenamefont {Skordis}}]{Clifton:2011jh}%
  \BibitemOpen
  \bibfield  {author} {\bibinfo {author} {\bibfnamefont {T.}~\bibnamefont
  {Clifton}}, \bibinfo {author} {\bibfnamefont {P.~G.}\ \bibnamefont
  {Ferreira}}, \bibinfo {author} {\bibfnamefont {A.}~\bibnamefont {Padilla}}, \
  and\ \bibinfo {author} {\bibfnamefont {C.}~\bibnamefont {Skordis}},\ }\href
  {\doibase 10.1016/j.physrep.2012.01.001} {\bibfield  {journal} {\bibinfo
  {journal} {Phys. Rept.}\ }\textbf {\bibinfo {volume} {513}},\ \bibinfo
  {pages} {1} (\bibinfo {year} {2012})},\ \Eprint
  {http://arxiv.org/abs/1106.2476} {arXiv:1106.2476 [astro-ph.CO]} \BibitemShut
  {NoStop}%
\bibitem [{\citenamefont {Nojiri}\ \emph {et~al.}(2017)\citenamefont {Nojiri},
  \citenamefont {Odintsov},\ and\ \citenamefont {Oikonomou}}]{Nojiri:2017ncd}%
  \BibitemOpen
  \bibfield  {author} {\bibinfo {author} {\bibfnamefont {S.}~\bibnamefont
  {Nojiri}}, \bibinfo {author} {\bibfnamefont {S.~D.}\ \bibnamefont
  {Odintsov}}, \ and\ \bibinfo {author} {\bibfnamefont {V.~K.}\ \bibnamefont
  {Oikonomou}},\ }\href {\doibase 10.1016/j.physrep.2017.06.001} {\bibfield
  {journal} {\bibinfo  {journal} {Phys. Rept.}\ }\textbf {\bibinfo {volume}
  {692}},\ \bibinfo {pages} {1} (\bibinfo {year} {2017})},\ \Eprint
  {http://arxiv.org/abs/1705.11098} {arXiv:1705.11098 [gr-qc]} \BibitemShut
  {NoStop}%
\bibitem [{\citenamefont {{Olmo}}\ and\ \citenamefont
  {{Rubiera-Garcia}}(2012)}]{Olmo2012a}%
  \BibitemOpen
  \bibfield  {author} {\bibinfo {author} {\bibfnamefont {G.~J.}\ \bibnamefont
  {{Olmo}}}\ and\ \bibinfo {author} {\bibfnamefont {D.}~\bibnamefont
  {{Rubiera-Garcia}}},\ }\href {\doibase 10.1103/PhysRevD.86.044014} {\bibfield
   {journal} {\bibinfo  {journal} {\prd}\ }\textbf {\bibinfo {volume} {86}},\
  \bibinfo {eid} {044014} (\bibinfo {year} {2012})},\ \Eprint
  {http://arxiv.org/abs/1207.6004} {arXiv:1207.6004 [gr-qc]} \BibitemShut
  {NoStop}%
\bibitem [{\citenamefont {{Olmo}}\ and\ \citenamefont
  {{Rubiera-Garc{\'\i}a}}(2012)}]{Olmo2012b}%
  \BibitemOpen
  \bibfield  {author} {\bibinfo {author} {\bibfnamefont {G.~J.}\ \bibnamefont
  {{Olmo}}}\ and\ \bibinfo {author} {\bibfnamefont {D.}~\bibnamefont
  {{Rubiera-Garc{\'\i}a}}},\ }\href {\doibase 10.1142/S0218271812500678}
  {\bibfield  {journal} {\bibinfo  {journal} {International Journal of Modern
  Physics D}\ }\textbf {\bibinfo {volume} {21}},\ \bibinfo {eid} {1250067}
  (\bibinfo {year} {2012})},\ \Eprint {http://arxiv.org/abs/1207.4303}
  {arXiv:1207.4303 [gr-qc]} \BibitemShut {NoStop}%
\bibitem [{\citenamefont {{Olmo}}\ and\ \citenamefont
  {{Rubiera-Garcia}}(2012)}]{Olmo2012c}%
  \BibitemOpen
  \bibfield  {author} {\bibinfo {author} {\bibfnamefont {G.~J.}\ \bibnamefont
  {{Olmo}}}\ and\ \bibinfo {author} {\bibfnamefont {D.}~\bibnamefont
  {{Rubiera-Garcia}}},\ }\href {\doibase 10.1140/epjc/s10052-012-2098-7}
  {\bibfield  {journal} {\bibinfo  {journal} {European Physical Journal C}\
  }\textbf {\bibinfo {volume} {72}},\ \bibinfo {eid} {2098} (\bibinfo {year}
  {2012})},\ \Eprint {http://arxiv.org/abs/1112.0475} {arXiv:1112.0475 [gr-qc]}
  \BibitemShut {NoStop}%
\bibitem [{\citenamefont {Olmo}\ \emph {et~al.}(2015)\citenamefont {Olmo},
  \citenamefont {Rubiera-Garcia},\ and\ \citenamefont
  {Sanchez-Puente}}]{Olmo:2015bya}%
  \BibitemOpen
  \bibfield  {author} {\bibinfo {author} {\bibfnamefont {G.~J.}\ \bibnamefont
  {Olmo}}, \bibinfo {author} {\bibfnamefont {D.}~\bibnamefont
  {Rubiera-Garcia}}, \ and\ \bibinfo {author} {\bibfnamefont {A.}~\bibnamefont
  {Sanchez-Puente}},\ }\href {\doibase 10.1103/PhysRevD.92.044047} {\bibfield
  {journal} {\bibinfo  {journal} {Phys. Rev. D}\ }\textbf {\bibinfo {volume}
  {92}},\ \bibinfo {pages} {044047} (\bibinfo {year} {2015})},\ \Eprint
  {http://arxiv.org/abs/1508.03272} {arXiv:1508.03272 [hep-th]} \BibitemShut
  {NoStop}%
\bibitem [{\citenamefont {Olmo}\ and\ \citenamefont
  {Rubiera-Garcia}(2023)}]{Olmo2022}%
  \BibitemOpen
  \bibfield  {author} {\bibinfo {author} {\bibfnamefont {G.~J.}\ \bibnamefont
  {Olmo}}\ and\ \bibinfo {author} {\bibfnamefont {D.}~\bibnamefont
  {Rubiera-Garcia}},\ }\enquote {\bibinfo {title} {Regular black holes in
  palatini gravity},}\ in\ \href {\doibase 10.1007/978-981-99-1596-5_6} {\emph
  {\bibinfo {booktitle} {Regular Black Holes: Towards a New Paradigm of
  Gravitational Collapse}}},\ \bibinfo {editor} {edited by\ \bibinfo {editor}
  {\bibfnamefont {C.}~\bibnamefont {Bambi}}}\ (\bibinfo  {publisher} {Springer
  Nature Singapore},\ \bibinfo {address} {Singapore},\ \bibinfo {year} {2023})\
  pp.\ \bibinfo {pages} {185--233}\BibitemShut {NoStop}%
\bibitem [{\citenamefont {Olmo}(2011)}]{Olmo:2011uz}%
  \BibitemOpen
  \bibfield  {author} {\bibinfo {author} {\bibfnamefont {G.~J.}\ \bibnamefont
  {Olmo}},\ }\href {\doibase 10.1142/S0218271811018925} {\bibfield  {journal}
  {\bibinfo  {journal} {Int. J. Mod. Phys. D}\ }\textbf {\bibinfo {volume}
  {20}},\ \bibinfo {pages} {413} (\bibinfo {year} {2011})},\ \Eprint
  {http://arxiv.org/abs/1101.3864} {arXiv:1101.3864 [gr-qc]} \BibitemShut
  {NoStop}%
\bibitem [{\citenamefont {Afonso}\ \emph
  {et~al.}(2018{\natexlab{a}})\citenamefont {Afonso}, \citenamefont {Olmo},\
  and\ \citenamefont {Rubiera-Garcia}}]{Afonso:2018bpv}%
  \BibitemOpen
  \bibfield  {author} {\bibinfo {author} {\bibfnamefont {V.~I.}\ \bibnamefont
  {Afonso}}, \bibinfo {author} {\bibfnamefont {G.~J.}\ \bibnamefont {Olmo}}, \
  and\ \bibinfo {author} {\bibfnamefont {D.}~\bibnamefont {Rubiera-Garcia}},\
  }\href {\doibase 10.1103/PhysRevD.97.021503} {\bibfield  {journal} {\bibinfo
  {journal} {Phys. Rev. D}\ }\textbf {\bibinfo {volume} {97}},\ \bibinfo
  {pages} {021503} (\bibinfo {year} {2018}{\natexlab{a}})},\ \Eprint
  {http://arxiv.org/abs/1801.10406} {arXiv:1801.10406 [gr-qc]} \BibitemShut
  {NoStop}%
\bibitem [{\citenamefont {Beltr\'an~Jim\'enez}\ and\ \citenamefont
  {Delhom}(2019)}]{BeltranJimenez:2019acz}%
  \BibitemOpen
  \bibfield  {author} {\bibinfo {author} {\bibfnamefont {J.}~\bibnamefont
  {Beltr\'an~Jim\'enez}}\ and\ \bibinfo {author} {\bibfnamefont
  {A.}~\bibnamefont {Delhom}},\ }\href {\doibase
  10.1140/epjc/s10052-019-7149-x} {\bibfield  {journal} {\bibinfo  {journal}
  {Eur. Phys. J. C}\ }\textbf {\bibinfo {volume} {79}},\ \bibinfo {pages} {656}
  (\bibinfo {year} {2019})},\ \Eprint {http://arxiv.org/abs/1901.08988}
  {arXiv:1901.08988 [gr-qc]} \BibitemShut {NoStop}%
\bibitem [{\citenamefont {Parker}\ and\ \citenamefont
  {Toms}(2009)}]{Parker2009}%
  \BibitemOpen
  \bibfield  {author} {\bibinfo {author} {\bibfnamefont {L.~E.}\ \bibnamefont
  {Parker}}\ and\ \bibinfo {author} {\bibfnamefont {D.}~\bibnamefont {Toms}},\
  }\href {\doibase 10.1017/CBO9780511813924} {\emph {\bibinfo {title} {{Quantum
  Field Theory in Curved Spacetime}: {Quantized Field and Gravity}}}},\
  Cambridge Monographs on Mathematical Physics\ (\bibinfo  {publisher}
  {Cambridge University Press},\ \bibinfo {year} {2009})\BibitemShut {NoStop}%
\bibitem [{\citenamefont {Afonso}\ \emph
  {et~al.}(2018{\natexlab{b}})\citenamefont {Afonso}, \citenamefont {Olmo},
  \citenamefont {Orazi},\ and\ \citenamefont
  {Rubiera-Garcia}}]{Afonso:2018mxn}%
  \BibitemOpen
  \bibfield  {author} {\bibinfo {author} {\bibfnamefont {V.~I.}\ \bibnamefont
  {Afonso}}, \bibinfo {author} {\bibfnamefont {G.~J.}\ \bibnamefont {Olmo}},
  \bibinfo {author} {\bibfnamefont {E.}~\bibnamefont {Orazi}}, \ and\ \bibinfo
  {author} {\bibfnamefont {D.}~\bibnamefont {Rubiera-Garcia}},\ }\href
  {\doibase 10.1140/epjc/s10052-018-6356-1} {\bibfield  {journal} {\bibinfo
  {journal} {Eur. Phys. J. C}\ }\textbf {\bibinfo {volume} {78}},\ \bibinfo
  {pages} {866} (\bibinfo {year} {2018}{\natexlab{b}})},\ \Eprint
  {http://arxiv.org/abs/1807.06385} {arXiv:1807.06385 [gr-qc]} \BibitemShut
  {NoStop}%
\bibitem [{\citenamefont {Olmo}\ \emph {et~al.}(2014)\citenamefont {Olmo},
  \citenamefont {Rubiera-Garcia},\ and\ \citenamefont
  {Sanchis-Alepuz}}]{Olmo:2013gqa}%
  \BibitemOpen
  \bibfield  {author} {\bibinfo {author} {\bibfnamefont {G.~J.}\ \bibnamefont
  {Olmo}}, \bibinfo {author} {\bibfnamefont {D.}~\bibnamefont
  {Rubiera-Garcia}}, \ and\ \bibinfo {author} {\bibfnamefont {H.}~\bibnamefont
  {Sanchis-Alepuz}},\ }\href {\doibase 10.1140/epjc/s10052-014-2804-8}
  {\bibfield  {journal} {\bibinfo  {journal} {Eur. Phys. J. C}\ }\textbf
  {\bibinfo {volume} {74}},\ \bibinfo {pages} {2804} (\bibinfo {year}
  {2014})},\ \Eprint {http://arxiv.org/abs/1311.0815} {arXiv:1311.0815
  [hep-th]} \BibitemShut {NoStop}%
\bibitem [{\citenamefont {{Visser}}(1995)}]{Visser1995}%
  \BibitemOpen
  \bibfield  {author} {\bibinfo {author} {\bibfnamefont {M.}~\bibnamefont
  {{Visser}}},\ }\href@noop {} {\emph {\bibinfo {title} {{Lorentzian wormholes.
  From Einstein to Hawking}}}}\ (\bibinfo {year} {1995})\BibitemShut {NoStop}%
\bibitem [{\citenamefont {{Stephani}}\ \emph {et~al.}(2009)\citenamefont
  {{Stephani}}, \citenamefont {{Kramer}}, \citenamefont {{MacCallum}},
  \citenamefont {{Hoenselaers}},\ and\ \citenamefont {{Herlt}}}]{Stephani2009}%
  \BibitemOpen
  \bibfield  {author} {\bibinfo {author} {\bibfnamefont {H.}~\bibnamefont
  {{Stephani}}}, \bibinfo {author} {\bibfnamefont {D.}~\bibnamefont
  {{Kramer}}}, \bibinfo {author} {\bibfnamefont {M.}~\bibnamefont
  {{MacCallum}}}, \bibinfo {author} {\bibfnamefont {C.}~\bibnamefont
  {{Hoenselaers}}}, \ and\ \bibinfo {author} {\bibfnamefont {E.}~\bibnamefont
  {{Herlt}}},\ }\href {\doibase 10.1017/CBO9780511535185} {\emph {\bibinfo
  {title} {{Exact Solutions of Einstein's Field Equations}}}}\ (\bibinfo {year}
  {2009})\BibitemShut {NoStop}%
\bibitem [{\citenamefont {Olmo}\ \emph {et~al.}(2016)\citenamefont {Olmo},
  \citenamefont {Rubiera-Garcia},\ and\ \citenamefont
  {Sanchez-Puente}}]{Olmo:2016fuc}%
  \BibitemOpen
  \bibfield  {author} {\bibinfo {author} {\bibfnamefont {G.~J.}\ \bibnamefont
  {Olmo}}, \bibinfo {author} {\bibfnamefont {D.}~\bibnamefont
  {Rubiera-Garcia}}, \ and\ \bibinfo {author} {\bibfnamefont {A.}~\bibnamefont
  {Sanchez-Puente}},\ }\href {\doibase 10.1088/0264-9381/33/11/115007}
  {\bibfield  {journal} {\bibinfo  {journal} {Class. Quant. Grav.}\ }\textbf
  {\bibinfo {volume} {33}},\ \bibinfo {pages} {115007} (\bibinfo {year}
  {2016})},\ \Eprint {http://arxiv.org/abs/1602.01798} {arXiv:1602.01798
  [hep-th]} \BibitemShut {NoStop}%
\bibitem [{\citenamefont {{Hawking}}(1971)}]{Hawking1971}%
  \BibitemOpen
  \bibfield  {author} {\bibinfo {author} {\bibfnamefont {S.}~\bibnamefont
  {{Hawking}}},\ }\href {\doibase 10.1093/mnras/152.1.75} {\bibfield  {journal}
  {\bibinfo  {journal} {\mnras}\ }\textbf {\bibinfo {volume} {152}},\ \bibinfo
  {pages} {75} (\bibinfo {year} {1971})}\BibitemShut {NoStop}%
\bibitem [{\citenamefont {Poisson}\ and\ \citenamefont
  {Will}(2014)}]{Poisson2014}%
  \BibitemOpen
  \bibfield  {author} {\bibinfo {author} {\bibfnamefont {E.}~\bibnamefont
  {Poisson}}\ and\ \bibinfo {author} {\bibfnamefont {C.~M.}\ \bibnamefont
  {Will}},\ }\href@noop {} {\emph {\bibinfo {title} {Gravity: Newtonian,
  post-newtonian, relativistic}}}\ (\bibinfo  {publisher} {Cambridge University
  Press},\ \bibinfo {year} {2014})\BibitemShut {NoStop}%
\bibitem [{\citenamefont {{Will}}(1993)}]{Will1993}%
  \BibitemOpen
  \bibfield  {author} {\bibinfo {author} {\bibfnamefont {C.~M.}\ \bibnamefont
  {{Will}}},\ }\href@noop {} {\emph {\bibinfo {title} {{Theory and Experiment
  in Gravitational Physics}}}}\ (\bibinfo {year} {1993})\BibitemShut {NoStop}%
\bibitem [{\citenamefont {{De Martino}}\ \emph {et~al.}(2021)\citenamefont {{De
  Martino}}, \citenamefont {{della Monica}},\ and\ \citenamefont {{De
  Laurentis}}}]{DeMartino2021}%
  \BibitemOpen
  \bibfield  {author} {\bibinfo {author} {\bibfnamefont {I.}~\bibnamefont {{De
  Martino}}}, \bibinfo {author} {\bibfnamefont {R.}~\bibnamefont {{della
  Monica}}}, \ and\ \bibinfo {author} {\bibfnamefont {M.}~\bibnamefont {{De
  Laurentis}}},\ }\href {\doibase 10.1103/PhysRevD.104.L101502} {\bibfield
  {journal} {\bibinfo  {journal} {\prd}\ }\textbf {\bibinfo {volume} {104}},\
  \bibinfo {eid} {L101502} (\bibinfo {year} {2021})},\ \Eprint
  {http://arxiv.org/abs/2106.06821} {arXiv:2106.06821 [gr-qc]} \BibitemShut
  {NoStop}%
\bibitem [{\citenamefont {{Della Monica}}\ \emph {et~al.}(2022)\citenamefont
  {{Della Monica}}, \citenamefont {{de Martino}},\ and\ \citenamefont {{de
  Laurentis}}}]{DellaMonica2022a}%
  \BibitemOpen
  \bibfield  {author} {\bibinfo {author} {\bibfnamefont {R.}~\bibnamefont
  {{Della Monica}}}, \bibinfo {author} {\bibfnamefont {I.}~\bibnamefont {{de
  Martino}}}, \ and\ \bibinfo {author} {\bibfnamefont {M.}~\bibnamefont {{de
  Laurentis}}},\ }\href {\doibase 10.1093/mnras/stab3727} {\bibfield  {journal}
  {\bibinfo  {journal} {\mnras}\ }\textbf {\bibinfo {volume} {510}},\ \bibinfo
  {pages} {4757} (\bibinfo {year} {2022})},\ \Eprint
  {http://arxiv.org/abs/2105.12687} {arXiv:2105.12687 [gr-qc]} \BibitemShut
  {NoStop}%
\bibitem [{\citenamefont {{Della Monica}}\ and\ \citenamefont {{de
  Martino}}(2022)}]{DellaMonica2022b}%
  \BibitemOpen
  \bibfield  {author} {\bibinfo {author} {\bibfnamefont {R.}~\bibnamefont
  {{Della Monica}}}\ and\ \bibinfo {author} {\bibfnamefont {I.}~\bibnamefont
  {{de Martino}}},\ }\href {\doibase 10.1088/1475-7516/2022/03/007} {\bibfield
  {journal} {\bibinfo  {journal} {\jcap}\ }\textbf {\bibinfo {volume} {2022}},\
  \bibinfo {eid} {007} (\bibinfo {year} {2022})},\ \Eprint
  {http://arxiv.org/abs/2112.01888} {arXiv:2112.01888 [astro-ph.GA]}
  \BibitemShut {NoStop}%
\bibitem [{\citenamefont {{Della Monica}}\ and\ \citenamefont {{de
  Martino}}(2023{\natexlab{a}})}]{DellaMonica2022c}%
  \BibitemOpen
  \bibfield  {author} {\bibinfo {author} {\bibfnamefont {R.}~\bibnamefont
  {{Della Monica}}}\ and\ \bibinfo {author} {\bibfnamefont {I.}~\bibnamefont
  {{de Martino}}},\ }\href {\doibase 10.1051/0004-6361/202245150} {\bibfield
  {journal} {\bibinfo  {journal} {\aap}\ }\textbf {\bibinfo {volume} {670}},\
  \bibinfo {eid} {L4} (\bibinfo {year} {2023}{\natexlab{a}})},\ \Eprint
  {http://arxiv.org/abs/2206.03980} {arXiv:2206.03980 [gr-qc]} \BibitemShut
  {NoStop}%
\bibitem [{\citenamefont {{Della Monica}}\ \emph
  {et~al.}(2023{\natexlab{a}})\citenamefont {{Della Monica}}, \citenamefont
  {{de Martino}}, \citenamefont {{Vernieri}},\ and\ \citenamefont {{de
  Laurentis}}}]{DellaMonica2023a}%
  \BibitemOpen
  \bibfield  {author} {\bibinfo {author} {\bibfnamefont {R.}~\bibnamefont
  {{Della Monica}}}, \bibinfo {author} {\bibfnamefont {I.}~\bibnamefont {{de
  Martino}}}, \bibinfo {author} {\bibfnamefont {D.}~\bibnamefont {{Vernieri}}},
  \ and\ \bibinfo {author} {\bibfnamefont {M.}~\bibnamefont {{de Laurentis}}},\
  }\href {\doibase 10.1093/mnras/stac3648} {\bibfield  {journal} {\bibinfo
  {journal} {\mnras}\ }\textbf {\bibinfo {volume} {519}},\ \bibinfo {pages}
  {1981} (\bibinfo {year} {2023}{\natexlab{a}})},\ \Eprint
  {http://arxiv.org/abs/2212.05082} {arXiv:2212.05082 [gr-qc]} \BibitemShut
  {NoStop}%
\bibitem [{\citenamefont {{GRAVITY Collaboration}}\ and\ \citenamefont
  {{Abuter}}(2018)}]{2018A&A...618L..10G}%
  \BibitemOpen
  \bibfield  {author} {\bibinfo {author} {\bibnamefont {{GRAVITY
  Collaboration}}}\ and\ \bibinfo {author} {\bibfnamefont {R.~{\it et al.}.}\
  \bibnamefont {{Abuter}}},\ }\href {\doibase 10.1051/0004-6361/201834294}
  {\bibfield  {journal} {\bibinfo  {journal} {\aap}\ }\textbf {\bibinfo
  {volume} {618}},\ \bibinfo {eid} {L10} (\bibinfo {year} {2018})},\ \Eprint
  {http://arxiv.org/abs/1810.12641} {arXiv:1810.12641 [astro-ph.GA]}
  \BibitemShut {NoStop}%
\bibitem [{\citenamefont {{Gillessen}}(2017)}]{Gillessen2017}%
  \BibitemOpen
  \bibfield  {author} {\bibinfo {author} {\bibfnamefont {S.~{\it et al.}.}\
  \bibnamefont {{Gillessen}}},\ }\href {\doibase 10.3847/1538-4357/aa5c41}
  {\bibfield  {journal} {\bibinfo  {journal} {\apj}\ }\textbf {\bibinfo
  {volume} {837}},\ \bibinfo {eid} {30} (\bibinfo {year} {2017})},\ \Eprint
  {http://arxiv.org/abs/1611.09144} {arXiv:1611.09144 [astro-ph.GA]}
  \BibitemShut {NoStop}%
\bibitem [{\citenamefont {Amorim}\ \emph {et~al.}(2019)\citenamefont {Amorim}
  \emph {et~al.}}]{GRAVITY:2019tuf}%
  \BibitemOpen
  \bibfield  {author} {\bibinfo {author} {\bibfnamefont {A.}~\bibnamefont
  {Amorim}} \emph {et~al.} (\bibinfo {collaboration} {GRAVITY}),\ }\href
  {\doibase 10.1093/mnras/stz2300} {\bibfield  {journal} {\bibinfo  {journal}
  {Mon. Not. Roy. Astron. Soc.}\ }\textbf {\bibinfo {volume} {489}},\ \bibinfo
  {pages} {4606} (\bibinfo {year} {2019})},\ \Eprint
  {http://arxiv.org/abs/1908.06681} {arXiv:1908.06681 [astro-ph.GA]}
  \BibitemShut {NoStop}%
\bibitem [{\citenamefont {{De Laurentis}}\ \emph {et~al.}(2023)\citenamefont
  {{De Laurentis}}, \citenamefont {{de Martino}},\ and\ \citenamefont {{Della
  Monica}}}]{DeLaurentis2023}%
  \BibitemOpen
  \bibfield  {author} {\bibinfo {author} {\bibfnamefont {M.}~\bibnamefont {{De
  Laurentis}}}, \bibinfo {author} {\bibfnamefont {I.}~\bibnamefont {{de
  Martino}}}, \ and\ \bibinfo {author} {\bibfnamefont {R.}~\bibnamefont {{Della
  Monica}}},\ }\href {\doibase 10.1088/1361-6633/ace91b} {\bibfield  {journal}
  {\bibinfo  {journal} {Reports on Progress in Physics}\ }\textbf {\bibinfo
  {volume} {86}},\ \bibinfo {eid} {104901} (\bibinfo {year} {2023})},\ \Eprint
  {http://arxiv.org/abs/2211.07008} {arXiv:2211.07008 [astro-ph.GA]}
  \BibitemShut {NoStop}%
\bibitem [{\citenamefont {Abuter}\ \emph
  {et~al.}(2020{\natexlab{b}})\citenamefont {Abuter} \emph
  {et~al.}}]{GRAVITY:2020xcu}%
  \BibitemOpen
  \bibfield  {author} {\bibinfo {author} {\bibfnamefont {R.}~\bibnamefont
  {Abuter}} \emph {et~al.} (\bibinfo {collaboration} {GRAVITY}),\ }\href
  {\doibase 10.1051/0004-6361/202037717} {\bibfield  {journal} {\bibinfo
  {journal} {Astron. Astrophys.}\ }\textbf {\bibinfo {volume} {638}},\ \bibinfo
  {pages} {A2} (\bibinfo {year} {2020}{\natexlab{b}})},\ \Eprint
  {http://arxiv.org/abs/2004.07185} {arXiv:2004.07185 [astro-ph.GA]}
  \BibitemShut {NoStop}%
\bibitem [{\citenamefont {{Cadoni}}\ \emph {et~al.}(2023)\citenamefont
  {{Cadoni}}, \citenamefont {{De Laurentis}}, \citenamefont {{De Martino}},
  \citenamefont {{Della Monica}}, \citenamefont {{Oi}},\ and\ \citenamefont
  {{Sanna}}}]{Cadoni2023}%
  \BibitemOpen
  \bibfield  {author} {\bibinfo {author} {\bibfnamefont {M.}~\bibnamefont
  {{Cadoni}}}, \bibinfo {author} {\bibfnamefont {M.}~\bibnamefont {{De
  Laurentis}}}, \bibinfo {author} {\bibfnamefont {I.}~\bibnamefont {{De
  Martino}}}, \bibinfo {author} {\bibfnamefont {R.}~\bibnamefont {{Della
  Monica}}}, \bibinfo {author} {\bibfnamefont {M.}~\bibnamefont {{Oi}}}, \ and\
  \bibinfo {author} {\bibfnamefont {A.~P.}\ \bibnamefont {{Sanna}}},\ }\href
  {\doibase 10.1103/PhysRevD.107.044038} {\bibfield  {journal} {\bibinfo
  {journal} {\prd}\ }\textbf {\bibinfo {volume} {107}},\ \bibinfo {eid}
  {044038} (\bibinfo {year} {2023})},\ \Eprint
  {http://arxiv.org/abs/2211.11585} {arXiv:2211.11585 [gr-qc]} \BibitemShut
  {NoStop}%
\bibitem [{\citenamefont {{\it et al.}}(2015)}]{Plewa2015}%
  \BibitemOpen
  \bibfield  {author} {\bibinfo {author} {\bibfnamefont {P.~M.~P.}\
  \bibnamefont {{\it et al.}}},\ }\href {\doibase 10.1093/mnras/stv1910}
  {\bibfield  {journal} {\bibinfo  {journal} {\mnras}\ }\textbf {\bibinfo
  {volume} {453}},\ \bibinfo {pages} {3234} (\bibinfo {year} {2015})},\ \Eprint
  {http://arxiv.org/abs/1509.01941} {arXiv:1509.01941 [astro-ph.GA]}
  \BibitemShut {NoStop}%
\bibitem [{\citenamefont {Gillessen}\ \emph {et~al.}(2009)\citenamefont
  {Gillessen}, \citenamefont {Eisenhauer}, \citenamefont {Fritz}, \citenamefont
  {Bartko}, \citenamefont {Dodds-Eden}, \citenamefont {Pfuhl}, \citenamefont
  {Ott},\ and\ \citenamefont {Genzel}}]{Gillessen2009a}%
  \BibitemOpen
  \bibfield  {author} {\bibinfo {author} {\bibfnamefont {S.}~\bibnamefont
  {Gillessen}}, \bibinfo {author} {\bibfnamefont {F.}~\bibnamefont
  {Eisenhauer}}, \bibinfo {author} {\bibfnamefont {T.~K.}\ \bibnamefont
  {Fritz}}, \bibinfo {author} {\bibfnamefont {H.}~\bibnamefont {Bartko}},
  \bibinfo {author} {\bibfnamefont {K.}~\bibnamefont {Dodds-Eden}}, \bibinfo
  {author} {\bibfnamefont {O.}~\bibnamefont {Pfuhl}}, \bibinfo {author}
  {\bibfnamefont {T.}~\bibnamefont {Ott}}, \ and\ \bibinfo {author}
  {\bibfnamefont {R.}~\bibnamefont {Genzel}},\ }\href {\doibase
  10.1088/0004-637X/707/2/L114} {\bibfield  {journal} {\bibinfo  {journal}
  {Astrophys. J. Lett.}\ }\textbf {\bibinfo {volume} {707}},\ \bibinfo {pages}
  {L114} (\bibinfo {year} {2009})},\ \Eprint {http://arxiv.org/abs/0910.3069}
  {arXiv:0910.3069 [astro-ph.GA]} \BibitemShut {NoStop}%
\bibitem [{\citenamefont {{Foreman-Mackey}}\ \emph {et~al.}(2013)\citenamefont
  {{Foreman-Mackey}}, \citenamefont {{Hogg}}, \citenamefont {{Lang}},\ and\
  \citenamefont {{Goodman}}}]{Foreman2013}%
  \BibitemOpen
  \bibfield  {author} {\bibinfo {author} {\bibfnamefont {D.}~\bibnamefont
  {{Foreman-Mackey}}}, \bibinfo {author} {\bibfnamefont {D.~W.}\ \bibnamefont
  {{Hogg}}}, \bibinfo {author} {\bibfnamefont {D.}~\bibnamefont {{Lang}}}, \
  and\ \bibinfo {author} {\bibfnamefont {J.}~\bibnamefont {{Goodman}}},\ }\href
  {\doibase 10.1086/670067} {\bibfield  {journal} {\bibinfo  {journal} {\pasp}\
  }\textbf {\bibinfo {volume} {125}},\ \bibinfo {pages} {306} (\bibinfo {year}
  {2013})},\ \Eprint {http://arxiv.org/abs/1202.3665} {arXiv:1202.3665
  [astro-ph.IM]} \BibitemShut {NoStop}%
\bibitem [{\citenamefont {{Gravity Collaboration}}\ and\ \citenamefont
  {{Abuter}}(2018)}]{GravityCollaboration2018a}%
  \BibitemOpen
  \bibfield  {author} {\bibinfo {author} {\bibnamefont {{Gravity
  Collaboration}}}\ and\ \bibinfo {author} {\bibfnamefont {R.~e.~a.}\
  \bibnamefont {{Abuter}}},\ }\href {\doibase 10.1051/0004-6361/201833718}
  {\bibfield  {journal} {\bibinfo  {journal} {\aap}\ }\textbf {\bibinfo
  {volume} {615}},\ \bibinfo {eid} {L15} (\bibinfo {year} {2018})},\ \Eprint
  {http://arxiv.org/abs/1807.09409} {arXiv:1807.09409 [astro-ph.GA]}
  \BibitemShut {NoStop}%
\bibitem [{\citenamefont {{Borka}}\ \emph {et~al.}(2012)\citenamefont
  {{Borka}}, \citenamefont {{Jovanovi{\'c}}}, \citenamefont {{Borka
  Jovanovi{\'c}}},\ and\ \citenamefont {{Zakharov}}}]{Borka2012}%
  \BibitemOpen
  \bibfield  {author} {\bibinfo {author} {\bibfnamefont {D.}~\bibnamefont
  {{Borka}}}, \bibinfo {author} {\bibfnamefont {P.}~\bibnamefont
  {{Jovanovi{\'c}}}}, \bibinfo {author} {\bibfnamefont {V.}~\bibnamefont
  {{Borka Jovanovi{\'c}}}}, \ and\ \bibinfo {author} {\bibfnamefont {A.~F.}\
  \bibnamefont {{Zakharov}}},\ }\href {\doibase 10.1103/PhysRevD.85.124004}
  {\bibfield  {journal} {\bibinfo  {journal} {\prd}\ }\textbf {\bibinfo
  {volume} {85}},\ \bibinfo {eid} {124004} (\bibinfo {year} {2012})},\ \Eprint
  {http://arxiv.org/abs/1206.0851} {arXiv:1206.0851 [astro-ph.CO]} \BibitemShut
  {NoStop}%
\bibitem [{\citenamefont {{Borka}}\ \emph {et~al.}(2013)\citenamefont
  {{Borka}}, \citenamefont {{Jovanovi{\'c}}}, \citenamefont {{Borka
  Jovanovi{\'c}}},\ and\ \citenamefont {{Zakharov}}}]{Borka2013}%
  \BibitemOpen
  \bibfield  {author} {\bibinfo {author} {\bibfnamefont {D.}~\bibnamefont
  {{Borka}}}, \bibinfo {author} {\bibfnamefont {P.}~\bibnamefont
  {{Jovanovi{\'c}}}}, \bibinfo {author} {\bibfnamefont {V.}~\bibnamefont
  {{Borka Jovanovi{\'c}}}}, \ and\ \bibinfo {author} {\bibfnamefont {A.~F.}\
  \bibnamefont {{Zakharov}}},\ }\href {\doibase 10.1088/1475-7516/2013/11/050}
  {\bibfield  {journal} {\bibinfo  {journal} {\jcap}\ }\textbf {\bibinfo
  {volume} {11}},\ \bibinfo {eid} {050} (\bibinfo {year} {2013})},\ \Eprint
  {http://arxiv.org/abs/1311.1404} {arXiv:1311.1404} \BibitemShut {NoStop}%
\bibitem [{\citenamefont {{D'Addio}}(2021)}]{DAddio2021}%
  \BibitemOpen
  \bibfield  {author} {\bibinfo {author} {\bibfnamefont {A.}~\bibnamefont
  {{D'Addio}}},\ }\href {\doibase 10.1016/j.dark.2021.100871} {\bibfield
  {journal} {\bibinfo  {journal} {Physics of the Dark Universe}\ }\textbf
  {\bibinfo {volume} {33}},\ \bibinfo {eid} {100871} (\bibinfo {year}
  {2021})}\BibitemShut {NoStop}%
\bibitem [{\citenamefont {{Della Monica}}\ \emph
  {et~al.}(2023{\natexlab{b}})\citenamefont {{Della Monica}}, \citenamefont
  {{de Martino}},\ and\ \citenamefont {{de Laurentis}}}]{DellaMonica2023b}%
  \BibitemOpen
  \bibfield  {author} {\bibinfo {author} {\bibfnamefont {R.}~\bibnamefont
  {{Della Monica}}}, \bibinfo {author} {\bibfnamefont {I.}~\bibnamefont {{de
  Martino}}}, \ and\ \bibinfo {author} {\bibfnamefont {M.}~\bibnamefont {{de
  Laurentis}}},\ }\href {\doibase 10.1093/mnras/stad579} {\bibfield  {journal}
  {\bibinfo  {journal} {\mnras}\ } (\bibinfo {year} {2023}{\natexlab{b}}),\
  10.1093/mnras/stad579},\ \Eprint {http://arxiv.org/abs/2302.12296}
  {arXiv:2302.12296 [gr-qc]} \BibitemShut {NoStop}%
\bibitem [{\citenamefont {{Della Monica}}\ and\ \citenamefont {{de
  Martino}}(2023{\natexlab{b}})}]{DellaMonica2023c}%
  \BibitemOpen
  \bibfield  {author} {\bibinfo {author} {\bibfnamefont {R.}~\bibnamefont
  {{Della Monica}}}\ and\ \bibinfo {author} {\bibfnamefont {I.}~\bibnamefont
  {{de Martino}}},\ }\href {\doibase 10.48550/arXiv.2305.10242} {\bibfield
  {journal} {\bibinfo  {journal} {arXiv e-prints}\ ,\ \bibinfo {eid}
  {arXiv:2305.10242}} (\bibinfo {year} {2023}{\natexlab{b}})},\ \Eprint
  {http://arxiv.org/abs/2305.10242} {arXiv:2305.10242 [gr-qc]} \BibitemShut
  {NoStop}%
\bibitem [{\citenamefont {Foschi}\ \emph {et~al.}(2023)\citenamefont {Foschi}
  \emph {et~al.}}]{GRAVITY:2023cjt}%
  \BibitemOpen
  \bibfield  {author} {\bibinfo {author} {\bibfnamefont {A.}~\bibnamefont
  {Foschi}} \emph {et~al.} (\bibinfo {collaboration} {GRAVITY}),\ }\href
  {\doibase 10.1093/mnras/stad1939} {\bibfield  {journal} {\bibinfo  {journal}
  {Mon. Not. Roy. Astron. Soc.}\ }\textbf {\bibinfo {volume} {524}},\ \bibinfo
  {pages} {1075} (\bibinfo {year} {2023})},\ \Eprint
  {http://arxiv.org/abs/2306.17215} {arXiv:2306.17215 [astro-ph.GA]}
  \BibitemShut {NoStop}%
\bibitem [{\citenamefont {Jim\'enez-Rosales}\ \emph {et~al.}(2020)\citenamefont
  {Jim\'enez-Rosales} \emph {et~al.}}]{GRAVITY:2020hwn}%
  \BibitemOpen
  \bibfield  {author} {\bibinfo {author} {\bibfnamefont {A.}~\bibnamefont
  {Jim\'enez-Rosales}} \emph {et~al.} (\bibinfo {collaboration} {GRAVITY}),\
  }\href {\doibase 10.1051/0004-6361/202038283} {\bibfield  {journal} {\bibinfo
   {journal} {Astron. Astrophys.}\ }\textbf {\bibinfo {volume} {643}},\
  \bibinfo {pages} {A56} (\bibinfo {year} {2020})},\ \Eprint
  {http://arxiv.org/abs/2009.01859} {arXiv:2009.01859 [astro-ph.HE]}
  \BibitemShut {NoStop}%
\bibitem [{\citenamefont {{Chen}}\ \emph {et~al.}(2019)\citenamefont {{Chen}},
  \citenamefont {{Bouhmadi-L{\'o}pez}},\ and\ \citenamefont
  {{Chen}}}]{Chen2019}%
  \BibitemOpen
  \bibfield  {author} {\bibinfo {author} {\bibfnamefont {C.-Y.}\ \bibnamefont
  {{Chen}}}, \bibinfo {author} {\bibfnamefont {M.}~\bibnamefont
  {{Bouhmadi-L{\'o}pez}}}, \ and\ \bibinfo {author} {\bibfnamefont
  {P.}~\bibnamefont {{Chen}}},\ }\href {\doibase
  10.1140/epjc/s10052-019-6585-y} {\bibfield  {journal} {\bibinfo  {journal}
  {European Physical Journal C}\ }\textbf {\bibinfo {volume} {79}},\ \bibinfo
  {eid} {63} (\bibinfo {year} {2019})},\ \Eprint
  {http://arxiv.org/abs/1811.12494} {arXiv:1811.12494 [gr-qc]} \BibitemShut
  {NoStop}%
\bibitem [{\citenamefont {{Rubiera-Garcia}}(2020)}]{Rubiera2020}%
  \BibitemOpen
  \bibfield  {author} {\bibinfo {author} {\bibfnamefont {D.}~\bibnamefont
  {{Rubiera-Garcia}}},\ }\href {\doibase 10.1142/S0218271820410072} {\bibfield
  {journal} {\bibinfo  {journal} {International Journal of Modern Physics D}\
  }\textbf {\bibinfo {volume} {29}},\ \bibinfo {eid} {2041007} (\bibinfo {year}
  {2020})},\ \Eprint {http://arxiv.org/abs/2004.00943} {arXiv:2004.00943
  [gr-qc]} \BibitemShut {NoStop}%
\bibitem [{\citenamefont {{Bahamonde}}\ and\ \citenamefont
  {{Valcarcel}}(2021)}]{Bahamonde2021}%
  \BibitemOpen
  \bibfield  {author} {\bibinfo {author} {\bibfnamefont {S.}~\bibnamefont
  {{Bahamonde}}}\ and\ \bibinfo {author} {\bibfnamefont {J.~G.}\ \bibnamefont
  {{Valcarcel}}},\ }\href {\doibase 10.1140/epjc/s10052-021-09275-6} {\bibfield
   {journal} {\bibinfo  {journal} {European Physical Journal C}\ }\textbf
  {\bibinfo {volume} {81}},\ \bibinfo {eid} {495} (\bibinfo {year} {2021})},\
  \Eprint {http://arxiv.org/abs/2103.12036} {arXiv:2103.12036 [gr-qc]}
  \BibitemShut {NoStop}%
\bibitem [{\citenamefont {{Lan}}\ \emph {et~al.}(2023)\citenamefont {{Lan}},
  \citenamefont {{Yang}}, \citenamefont {{Guo}},\ and\ \citenamefont
  {{Miao}}}]{Lan2023}%
  \BibitemOpen
  \bibfield  {author} {\bibinfo {author} {\bibfnamefont {C.}~\bibnamefont
  {{Lan}}}, \bibinfo {author} {\bibfnamefont {H.}~\bibnamefont {{Yang}}},
  \bibinfo {author} {\bibfnamefont {Y.}~\bibnamefont {{Guo}}}, \ and\ \bibinfo
  {author} {\bibfnamefont {Y.-G.}\ \bibnamefont {{Miao}}},\ }\href {\doibase
  10.1007/s10773-023-05454-1} {\bibfield  {journal} {\bibinfo  {journal}
  {International Journal of Theoretical Physics}\ }\textbf {\bibinfo {volume}
  {62}},\ \bibinfo {eid} {202} (\bibinfo {year} {2023})},\ \Eprint
  {http://arxiv.org/abs/2303.11696} {arXiv:2303.11696 [gr-qc]} \BibitemShut
  {NoStop}%
\end{thebibliography}%

\end{document}